\documentclass[%
reprint,
superscriptaddress,
nofootinbib,
amsmath,amssymb,
aps,
pre,
notitlepage,
showkeys]{revtex4-2}
\usepackage[colorlinks=true, urlcolor=blue, anchorcolor=blue, citecolor=blue,filecolor=blue,linkcolor=blue,menucolor=blue]{hyperref}
\usepackage{url}
\usepackage{graphicx}
\usepackage{lipsum}
\usepackage[export]{adjustbox}
\usepackage{mathtools}
\usepackage{gensymb}
\usepackage{csquotes}
\usepackage[T1]{fontenc}
\usepackage{amsmath}
\usepackage{amssymb}
\usepackage{stmaryrd}
\usepackage{fancyhdr}
\usepackage[utf8]{inputenc}
\usepackage{rotating}
\usepackage{comment}
\usepackage{url}
\usepackage{bbm}
\usepackage{soul}
\usepackage[normalem]{ulem}
\usepackage{tabularx,booktabs}
\usepackage{stmaryrd}
\newcolumntype{Y}{>{\centering\arraybackslash}X}

\usepackage{hyperref}  
\usepackage{subfigure}

\begin{document}
\title{Quantifying Broken Detailed Balance in Transcription}
\author{James Holehouse}
\email{jamesholehouse1@gmail.com}
\affiliation{The Santa Fe Institute, 1399 Hyde Park Road, Santa Fe, NM, 87501, USA}

\begin{abstract}
    For the canonical two-state model of transcription, we derive exact analytic expressions for the entropy production rate of transcription at steady state, and assess detailed balance breaking in transcription. Our analytics allow us to easily evaluate the entropy production rate of thousands of genes across seven datasets of two-state model parameters without needing to evaluate the entropy production rate from trajectory-based computation. A data-driven approach then exposes that most genes avoid parameter regimes associated with large entropy production rates, akin to a mesoscopic version of energy expenditure minimization. Importantly, we show that this is not a thermodynamic phenomenon, since the entropy production rate from the two state gene model provides only a weak bound on the housekeeping energy needed to power transcription. Finally, we show that cell-to-cell variability can make mRNA expression seem more or less irreversible than a ``representative cell'' would imply.
\end{abstract}

\maketitle




\section*{Introduction}

\noindent Cells are fundamentally non-equilibrium systems \cite{schrodinger2012life}. They metabolize free energy to drive processes such as transcription, translation, regulation, and replication—activities necessary for survival and reproduction. If free energy sources are not present, mRNA and protein production ceases. Eventually, the system relaxes toward equilibrium with its environment, and the cell ultimately dies. In a proliferating population, this non-equilibrium character is evident in the irreversibility of exponential growth \cite{england2013statistical,grandpre2025extremal,ocal2025two}. Yet, the extent to which the molecular processes underpinning this irreversibility violate detailed balance---and the degree to which reversibility is a determining factor in the dynamics of molecular machinery---remains an open question \cite{battle2016broken}.

In this paper, we take interest in understanding how the behaviors exhibited in the transcription of genes may be related to the ways in which those behaviors break \textit{time-reversal symmetry}---otherwise known as detailed balance. Fundamentally, transcription is a noisy process that is far-from-equilibrium \cite{zoller2022eukaryotic,wong2020gene,raj2006stochastic}. This noise arises not only from extrinsic factors that vary between cells but also from the random waiting times between molecular collisions among genes, mRNA, and protein molecules \cite{mcadams1997stochastic,elowitz2002stochastic,swain2002intrinsic,raj2008nature,VERHAGEN2025169202}. The most common models of transcription are phenomenological---namely, stochastic switching of the gene between active (on) and inactive (off) states, transcription occurring only when the gene is active, and the subsequent degradation of mRNA transcripts. Simple models are highly successful at capturing the dynamics of genes that have mRNA distributions that are noisier than a Poisson point process \cite{braichenko2021distinguishing}.  Under these conditions, gene expression is said to be bursty, being characterized by short on-times and long waits in between bursts of mRNA production. This ``bursty'' nature is studied both at the level of transcript distributions across population snapshots \cite{molina2013stimulus,bartman2019transcriptional,volteras2023global,fu2022quantifying,sukys2025cell} and in time for individual genes \cite{golding2005real,suter2011mammalian,lammers2020matter,pomp2024transcription}.


\begin{figure}[ht]
    \centering
    \includegraphics[width=0.5\textwidth]{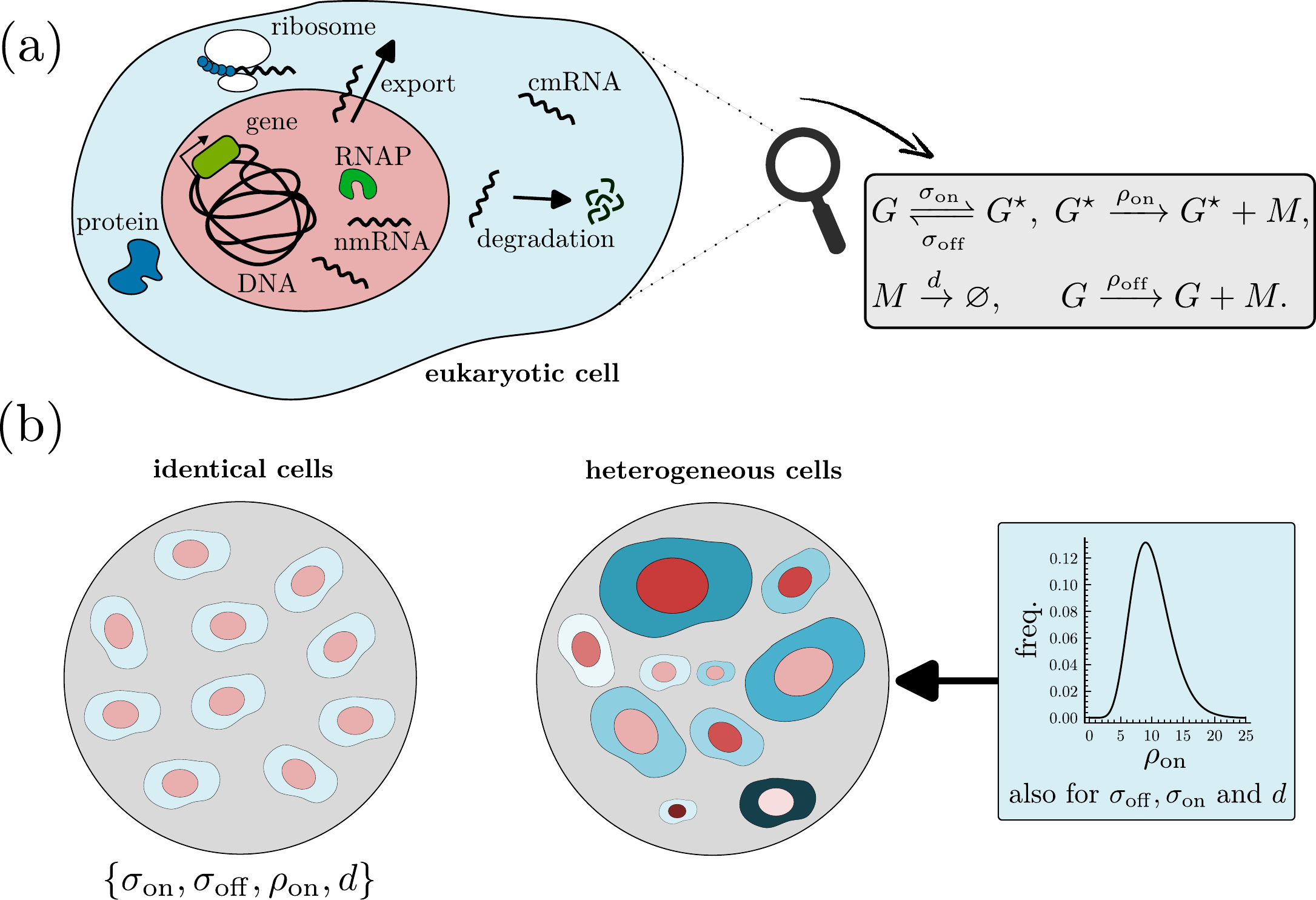}
    \caption{\textbf{Schematics of the models considered in this study}. (a) Illustration of a eukaryotic cell with the most important molecules and processes labeled. The gray box shows the model of transcription considered in this study. (b) Typical modeling and inference schemes assume populations of identical cells, and that most of cellular noise is intrinsic. A population of heterogeneous cells including varying features such as cell size or activator concentrations, leading to distributions of kinetic parameters over the populations \cite{lim2015quantitative,grima2023quantifying}.}
    \label{fig:schematic}
\end{figure}

Stochastic thermodynamic analyses aimed at quantifying the degree to which mRNA expression dynamics violate detailed balance are relatively uncommon \cite{tkavcik2008information,gama2020binary,boeger2022energetics,boeger2022kinetic,zoller2022eukaryotic}. In part, this is because the common models of transcription are phenomenological, and Markov state transitions do not correspond to individual energy barriers. For instance, in models of constitutive expression---where a gene is presumed to produce mRNA in a Poisson process---the underlying process is, in reality, composed of many elementary steps describing the stepping of RNA Pol II along a gene. The Poissonian steady state described by constitutive expression, i.e., $G\to G+M,\;M\to\varnothing$, although in detailed balance is not an equilibrium because the reservoirs containing the ATP to fuel the process are chemostatted \cite{qian2007phosphorylation}. One principled way to bridge this gap between macroscopic models and microscopic thermodynamics is to consider the entropy production of a coarse-grained model as a lower bound to the entropy produced by the fine-grained system \cite{egolf2000equilibrium,blythe2008reversibility,gomez2008lower,england2013statistical,bo2017multiple,jia2016model,skinner2021estimating}. However, such bounds may not always be tight \cite{ghosal2022inferring}, and in some cases can even be violated if care is not taken in how decay products and reactants are distinguished \cite{kolchinsky2024thermodynamic}.

Rather than asking what the thermodynamics of a coarse-grained process can reveal about a more fine-grained process, one can instead consider the reverse: \textit{how does mesoscopic or macroscopic irreversibility emerge from the irreversibility or more fine-grained behavior?}
For example, in \cite{lynn2021broken,lynn2022emergence}, Lynn \textit{et al.}~devise algorithms to explore how exposing human subjects to different tasks changes the level of broken detailed balance in the brain, making the case that violations of detailed balance are necessary for cognition. From this perspective, one can ask how the irreversibility of a proliferating population of cells may arise from the dynamics of individual genes, and how the reality of cell-to-cell variability may affect these interpretations. A major question this paper tackles is: \textit{how does the irreversible nature of cellular dynamics propagate down to the level of individual genes?}


In this work, we develop simple analytic tools to study the breaking of detailed balance in transcription, under the framework of a two-state model of gene expression. Our analytic expressions allow entropy production rates to be directly computed from a gene’s kinetic parameters, allowing us to avoid simulation-based methods of calculating dynamical reversibility at a trajectory level. We apply these tools to estimate entropy production across thousands of genes in mice, across seven datasets of inferred two-state model parameters.  Although the thermodynamic bounds provided by the telegraph model's entropy production rate are not tight, we show that real genes choose kinetic parameters that clearly avoid areas of high entropy production. Defining two statistical null models we show that similar properties are not as pronounced when transcriptional kinetic parameters are randomized. Finally, we assess analytically how cell-to-cell variability affects the irreversibility of gene expression at the population level, and the potential implications this has in determining where the cell may tolerate cell-to-cell variability.

\subsection*{Two-state model of transcription}\label{sec:two-state-model}

\noindent Although there is substantial evidence suggesting that transcriptional kinetics are more complex than a simple two-state Markovian gene model implies \cite{suter2011mammalian,bothma2014dynamic, rodriguez2019intrinsic, bartman2019transcriptional, cao2020stochastic, tunnacliffe2020transcriptional, shelansky2024single}, all widely accessible parameter inference datasets have relied on a two-state framework. In this model, key parameters such as the gene’s activation rate, inactivation rate, and transcription rate have been estimated across thousands of genes \cite{larsson2019genomic, ramskold2024single, sukys2025cell}. Generally, the two-state model is considered under the further assumption that a gene admits no ``leakiness'', and that a transcriptionally off-state cannot be transcribed---this model is known as the \textit{telegraph model} \cite{peccoud1995markovian,raj2006stochastic,iyer2009stochasticity}. Two-state gene models are similar to models in the physics literature that have studied the stochastic dynamics of systems with switching environments \cite{hufton2016intrinsic}.

A major advantage of the two-state model is its distinguishability: from population level mRNA sequencing of mRNA FISH experiments, its three kinetic parameters can be robustly inferred. In contrast, comprehensive databases of kinetic parameters inferred from models with more than two gene states do not yet exist, in part because of the difficulty in performing experiments that would allow for transitions between gene states to be distinguished on a the level of a single cell \cite{bartman2019transcriptional,wildner2023bayesian,eck2024single}. Other complications in parameter inference include the reality that transcriptional kinetics are dependent on factors that vary from cell-to-cell, and it has been concluded by some that mRNA transcript variability is determined by the properties of individual cells and not noise intrinsic to transcription and translation \cite{battich2015control,ham2020extrinsic,grima2023quantifying}.



The two-state model considered in this paper is given by the set of reactions,
\begin{align}\label{eq:modTM}
    G\xrightarrow{\rho_{\mathrm{off}}}G+M,\; G^\star\xrightarrow{\rho_{\mathrm{on}}}G^\star+M,\;G\xrightleftharpoons[\sigma_{\mathrm{off}}]{\sigma_{\mathrm{on}}} G^\star,\; M\xrightarrow{d}\varnothing.
\end{align}
$G$ and $G^\star$ represent the two different gene states with $n_G+n_{G^\star}=1$ (there is a single gene) and $M$ represents mRNA. This is also known as the \textit{leaky telegraph model} or \textit{leaky gene model} which was first solved in the steady state regime in \cite{ham2020exactly}. 
Standard narratives ascribe $\rho_{\mathrm{off}}=0$ such that $G$ is a true off state.  However, this assumption is not required, and we arbitrarily set $\rho_{\mathrm{on}}>\rho_{\mathrm{off}}$. It is well established that gene repression is not perfect, leading to basal expression---across the tree of life \cite{phan2015development,zacharioudakis2016novel,flouriot2020basal}. In practice, two-state models coarse-grain over a wide range of underlying molecular processes, and in the general case, the off state may simply exhibit a reduced transcription rate rather than none at all. This leads to a mesoscopic model of transcriptional dynamics that is dynamically reversible---also known as \textit{weak reversibility}---although the dynamics are generally not reversible in the thermodynamic sense. 
A schematic of this process is shown in Fig.~\ref{fig:schematic}(a). The turning on (off) of the gene state could be related to the binding (unbinding) of an abundant activating transcription factor (and \textit{vice versa} for a repressor), and the loss of $M$ is due to degradation.

\begin{figure}[ht]
    \includegraphics[width=.48\textwidth]{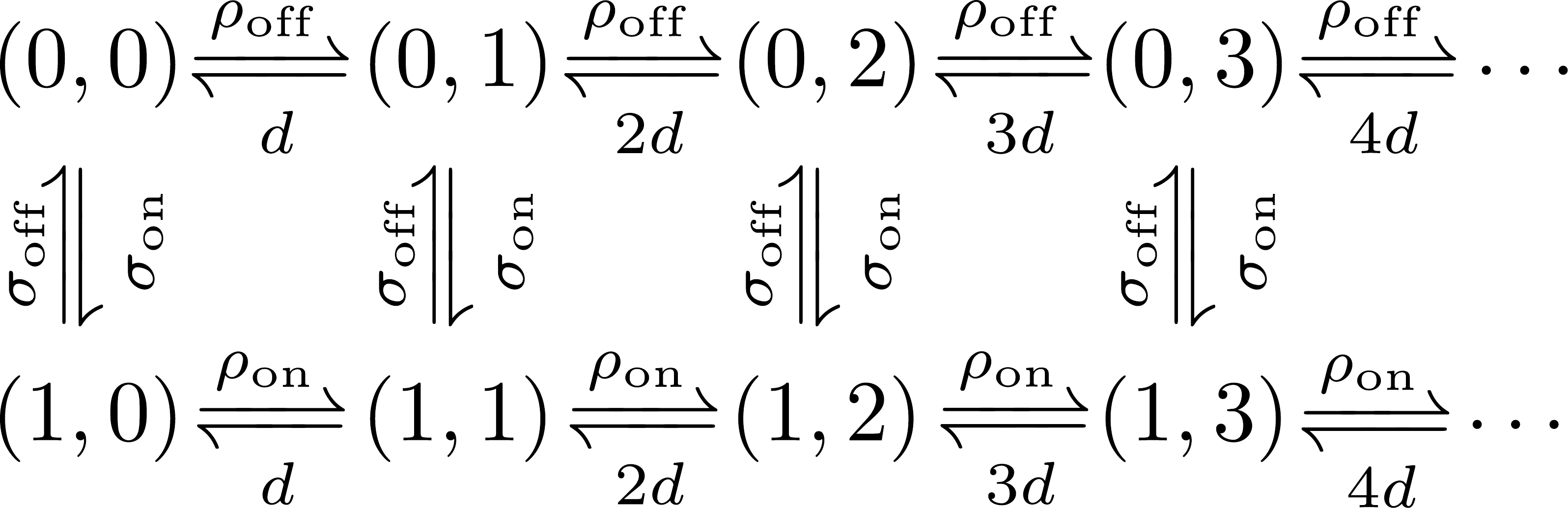}
    \caption{\textbf{Markov state diagram for the reaction scheme in Eq.~\eqref{eq:modTM} for states up to mRNA number $n=3$}. Arrow labels indicate the propensity at which transitions between each state occur. Since every reaction can occur in both directions, the telegraph model is a dynamically reversible non-equilibrium system. Generally, detailed balance is not satisfied.}
    \label{fig:markov-d-TM}
\end{figure}

Below, $\mathbf{x}=(n_G^\star,n)$ denotes the state vector describing the number of $G^\star$ and $M$ respectively (with the number of $G^\star$ being $1-n_G$), this set of reactions has the Markov state diagram shown in Fig.~\ref{fig:markov-d-TM}. A steady state that admits detailed balance is only guaranteed for one-dimensional reversible Markov chains, or for dynamically reversible reactions in closed vessels with specific parameter choices (see the final equation in \cite{van1976equilibrium} and generally \cite{van1992stochastic}).
The probabilities $P_0(n)$ and $P_1(n)$ denoting the probabilities of having states $(0,n)$ and $(1,n)$ are described by the master equations,
\begin{align}\label{eq:Meqs}
\begin{split}
    \partial_t P_0(n, t) =& \rho_{\mathrm{off}}  (\mathbb{E}^{-1}-1)P_0(n,t) + d(\mathbb{E}^{1}-1)nP_0(n,t) \\&+ \sigma_{\mathrm{off}} P_1(n,t) - \sigma_{\mathrm{on}} P_0(n,t), \\
    \partial_t P_1(n, t) =& \rho_{\mathrm{on}} (\mathbb{E}^{-1}-1)P_0(n,t) + d(\mathbb{E}^{1}-1)nP_1(n,t) \\&- \sigma_{\mathrm{off}}  P_1(n,t) + \sigma_{\mathrm{on}} P_0(n,t),
\end{split}
\end{align}
and $\mathbb{E}^x$ is the step operator that acts such that $\mathbb{E}^x f(n) = f(n+x)$ \cite{van1992stochastic}. One can then introduce the generating functions $G_0(z)=\sum_n z^n P_0(n)$ and $G_1(z)=\sum_n z^n P_1(n)$ and solve the resulting ODE at steady state when $\partial_tG_0=\partial_t G_1=0$ to find,
\begin{align}\nonumber
    G_0(z) &= \frac{\sigma_{\mathrm{off}}  e^{\rho_{\mathrm{off}} (z-1)/d}}{\Sigma}{}_1F_1\left(\frac{\sigma_{\mathrm{on}} }{d},1+\frac{\Sigma}{d};\frac{(z-1)\delta}{d}\right),\\\nonumber
    G_1(z) &= \frac{\sigma_{\mathrm{on}}  e^{\rho_{\mathrm{off}} (z-1)/d}}{\Sigma}{}_1F_1\left(\frac{d+\sigma_{\mathrm{on}} }{d},1+\frac{\Sigma}{d};\frac{(z-1)\delta}{d}\right),
\end{align}
using definitions $\Sigma=\sigma_{\mathrm{off}} +\sigma_{\mathrm{on}} $ and $\delta = \rho_{\mathrm{on}} -\rho_{\mathrm{off}} $ and where ${}_1F_1$ denotes the confluent hypergeometric function. The steady state probabilities $P_0(n)$ and $P_1(n)$ are then recovered from the series expansions of $G_0(z)$ and $G_1(z)$ about $z=0$.
Note that the sum $G=G_0+G_1$ can be used to calculate $P(n)=P_0(n)+P_1(n)$ which is given by,
\begin{align}\label{eq:Gtm}
    G(z) = e^{\rho_{\mathrm{off}} (z-1)/d}{}_1F_1\left(\frac{\sigma_{\mathrm{on}} }{d},\frac{\Sigma}{d};\frac{(z-1)\delta}{d}\right).
\end{align}
This solution has also been provided in \cite{ham2020exactly}, although $G_0(z)$ and $G_1(z)$ are not provided therein. For later use, one can show that in the ``bursty limit'' of $\{\rho_{\mathrm{on}},\sigma_{\mathrm{off}}\}\gg \{\sigma_{\mathrm{off}},\rho_{\mathrm{off}},d\}$ that the Fano factor ($\mathrm{FF}$) and coefficient of variation squared ($CV^2$) of mRNA molecule numbers are given by,
\begin{align}\label{eq:ff}
    \mathrm{FF} &= 1+\frac{B^2 \sigma_\mathrm{on}}{\rho_{\mathrm{off}}+B \sigma_{\mathrm{on}}},\\\label{eq:cv2}
    CV^2 &= \frac{d(\rho_{\mathrm{off}}+B(1+B)\sigma_{\mathrm{on}})}{(\rho_{\mathrm{off}}+B\sigma_{\mathrm{on}})^2}.
\end{align}
Note that these expressions are also calculable in the non-bursty regime, but the expressions are more complicated. Derivations of these expressions can be found in Sec.~\ref{sec:ffcv2}.

To understand the non-equilibrium nature of the two-state gene, one can use the equations for $G_0(z)$ and $G_1(z)$ to calculate the probability fluxes between the Markov states in Fig.~\ref{fig:markov-d-TM}. First, it is clear that if $\rho_{\mathrm{off}} =\rho_{\mathrm{on}} $ then the gene states cannot be distinguished and the generating function reduces to that of a Poisson process, since ${}_1F_1(a,b;0)=1$. Typically, in the context of bursty gene expression (where $\rho_{\mathrm{off}} \ll \rho_{\mathrm{on}}$) $\sigma_{\mathrm{on}}$ is known as the \textit{burst frequency} and $B=\rho_{\mathrm{on}} /\sigma_{\mathrm{off}} $ as the \textit{mean burst size}. Although there is no commonly accepted definition of how bursty a gene is, here we use two measures: (i) $\sigma_{\mathrm{off}}/\sigma_{\mathrm{on}}$, which is the ratio of timescale of off switching to on switching---larger $\sigma_{\mathrm{off}}/\sigma_{\mathrm{on}}$ implies a greater degree of burstiness because more time is spent in a transcriptionally impaired state; and (ii) $CV^2$, because bursting is often thought of as confining variability to the level of mRNA transcripts, and $CV^2$ is a direct measure of transcript variability.


\begin{figure}[ht]
    \includegraphics[width=.48\textwidth]{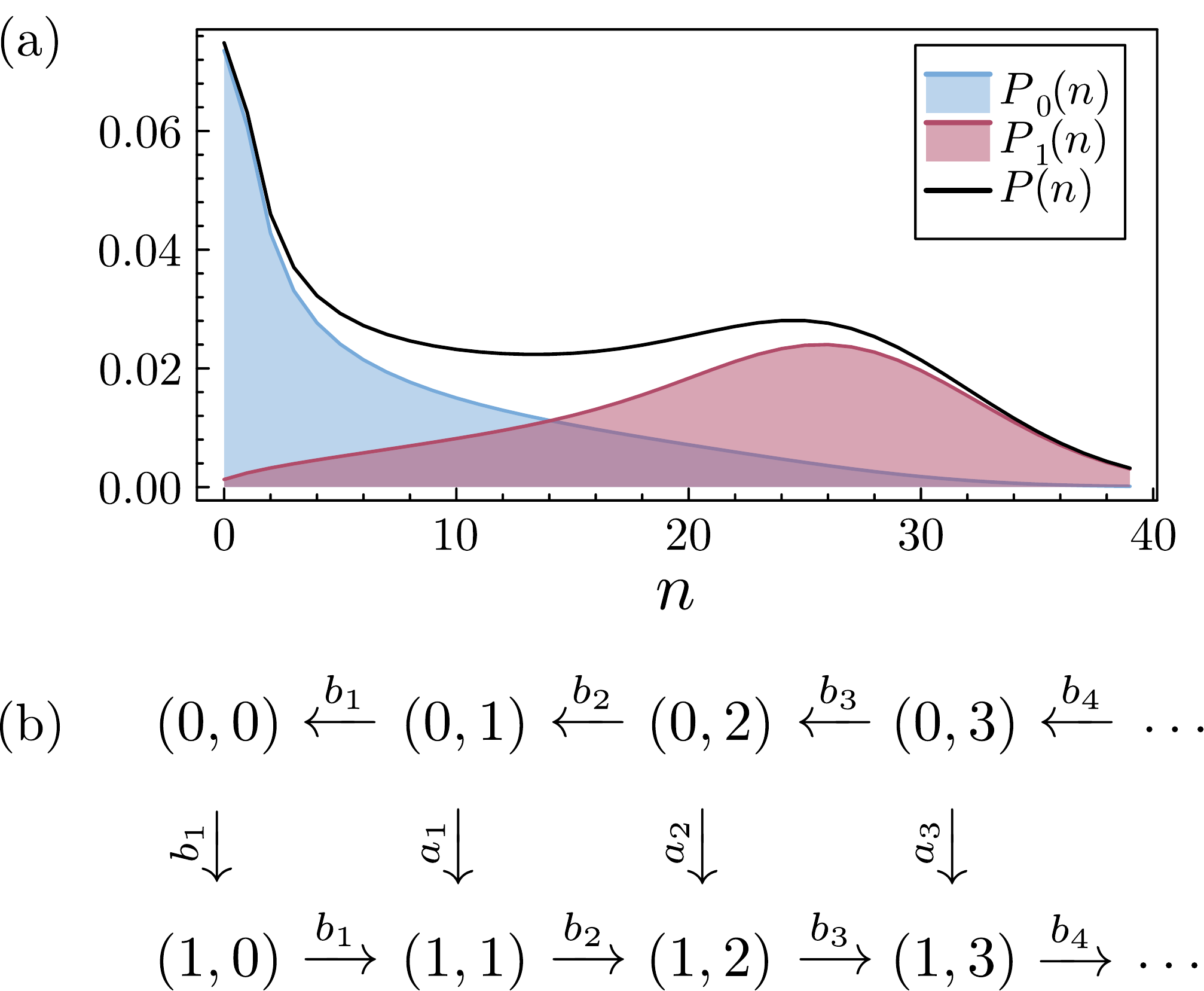}
    \caption{\textbf{Exploring the non-equilibrium steady state of the two-state model.} (a) Plots of $P_0(n)$, $P_1(n)$ and $P(n)$ calculated using the generating functions in the main text for parameters $\rho_{\mathrm{off}} = 1/3, \rho_{\mathrm{on}}=30, \sigma_{\mathrm{off}} = \sigma_{\mathrm{on}} = 1/2,$ and $d = 1$. (b) A diagram showing the flow of probability flux between states in the Markov diagram. Variables $b_n$ and $a_n$ represent probability fluxes, whose expressions are given in the main text. Here $b_n>0$, whereas the $a_n$ can be positive or negative.}
    \label{fig:fig2}
\end{figure}

Fig.~\ref{fig:fig2} shows a probability flux analysis on the two-state gene model. 
To get some intuition for the nature of the steady state one can look at the net probability flux between each pair of connected states,
shown in Fig.~\ref{fig:fig2}(b) \cite{schnakenberg1976network}. In words, there is a unidirectional flux of probability from right-to-left along the states $(0,n)\to(0,n-1)\,\forall \,n\in \{1,2,\ldots\}$ and from left-to-right along the states $(1,n-1)\to(1,n)\,\forall \,n\in \{1,2,\ldots\}$ which in both cases is of magnitude $b_n>0$. Between the two chains there are fluxes $a_n$ between states $(0,n)$ and $(1,n)$ which can be positive or negative, with the exception of $a_0 = b_1>0$ to conserve probability at the end of the chain.

An interesting observation is that if one ignores the state of the gene, \textit{on the level of the mRNA detailed balance is maintained}. This is also true if one were to lump together the mRNA states and only observe the gene state dynamics, since one can show that $-b_1=\sum_{i=1}^\infty a_i$. This finding, that lumping gene states gives marginal dynamics satisfying detailed balance, aligns with other recent findings that non-equilibrium systems with hidden states can ``pretend'' to obey detailed balance \cite{skinner2021estimating}. \textit{Marginal detailed balance} on the mRNA number means that the condition,
\begin{align}\label{eq:EffDBCond}
    b_n\equiv n d P_0(n)-\rho_{\mathrm{off}} P_0(n-1) = \rho_{\mathrm{on}} P_1(n-1)-n d P_1(n)
\end{align}
must be true, which can be shown directly from the generating functions above. This marginal detailed balance condition is hinted to by the fact that the eigenvalues governing the relaxation to the steady state distribution are \textit{real} \cite{van1992stochastic,tauber2014critical}, given by the two sets $\lambda^1_n = n d,\; n\in \{0,1,2,\ldots\}$ and $\lambda^2_n =n d + \sigma_{\mathrm{on}}+ \sigma_{\mathrm{off}},\; n\in \{0,1,2,\ldots\}$,
which is shown in Section \ref{sec:TMsol}. Note that, although it is true that a system satisfying detailed balance will have a master operator with real eigenvalues, the converse is generally not true. This two-state gene model is an example of this fact. We explore further implications of marginal detailed balance in Section \ref{sec:red-mods}. 

\subsection*{Entropy production of transcription in a two-state gene}\label{sec:EPR}

\noindent To investigate the degree to which an individual gene---comprising kinetic parameters $\sigma_{\mathrm{on}}, \sigma_{\mathrm{off}}, \rho_{\mathrm{on}}, \rho_{\mathrm{off}}$ and $d$---breaks detailed balance, we can utilize a quantity known as the \textit{entropy production rate} (EPR) \cite{peliti2021stochastic,seifert2012stochastic,boeger2022kinetic}. The EPR has a couple of very useful properties. First, the EPR tells us the degree to which a process is irreversible. For a system in detailed balance, the EPR is zero because every forward transition has an equal reverse probability flux---this is the requirement for time-reversal symmetry. Zero is the minimum value of the EPR. For a dynamically irreversible system, the EPR is infinite because not every forward transition has a reverse transition---this is the requirement to have forward trajectories that could never be seen in reverse. Second, for a microscopic system made up of elementary reactions, the EPR is proportional to the amount of free energy needed to power the system in its non-equilibrium dynamics \cite{ge2013dissipation,yoshimura2023housekeeping}. The aims of this section are: (i) to find an analytic formula for the EPR of a transcribing gene; (ii) to apply this formula across several kinetic parameter datasets for genes in mouse cells; (iii) to understand the degree to which this EPR gives a thermodynamic lower bound on the microscopic EPR; and (iv) to discover, from a data-driven approach, whether reversibility is a desirable property in the dynamics of mRNA expression.

Following standard methods \cite{peliti2021stochastic,boeger2022kinetic}, the EPR is given by (in natural units),
\begin{align}\label{eq:epr}
    \dot{s}_{\mathrm{mes}} = \frac{1}{2}\sum_{\mathbf{x},\mathbf{x'}} J(\mathbf{x'}\to\mathbf{x})\ln\left(  \frac{w(\mathbf{x'}\to\mathbf{x})P(\mathbf{x}',t)}{w(\mathbf{x}\to\mathbf{x'})P(\mathbf{x},t)}\right),
\end{align}
in which $J(\mathbf{x'}\to\mathbf{x})$ and $w(\mathbf{x'}\to\mathbf{x})$ are the respective net flux and propensity from $\mathbf{x}$ to $\mathbf{x'}$, and $P(\mathbf{x},t)$ is probability of having state $\mathbf{x}$ at time $t$. The subscript ``mes'' denotes that this refers to a mesoscopic EPR. 

The EPR in Eq.~\eqref{eq:epr} is often broken down into components of \textit{entropy flow rate}---entropy production due to interactions between the system and an external reservoir of energy or particles---
and \textit{the entropy change of the system}---relating to changes in the internal configurations of the system \cite{schnakenberg1976network}. However, we assume the system is at a steady state, meaning that the EPR is simply the negative of the entropy flow rate. It is equal to the \textit{negative} entropy flow rate due to the convention in stochastic thermodynamics that it is the system doing the work, and not having work done on it \cite{peliti2021stochastic}. Negative entropy flow therefore implies that the system is subject to influx from the reservoirs. The equation for the entropy flow rate, $\dot{s}_{e}$, is (in natural units),
\begin{align}\label{eq:epr_flow}
    \dot{s}_{e} \equiv \frac{1}{2}\sum_{\mathbf{x},\mathbf{x'}} J(\mathbf{x'}\to\mathbf{x})\ln\left(  \frac{w(\mathbf{x}\to\mathbf{x'})}{w(\mathbf{x'}\to\mathbf{x})}\right) = -\dot s_{\mathrm{mes}}.
\end{align}
For further details on the relationship between the entropy production rate and the entropy flow rate at steady state, please refer to Section \ref{sec:EPRcalc}, and references \cite{schnakenberg1976network,boeger2022energetics,kirchberg2023energy,esposito2010three}.

For the calculation of the two-state gene model's EPR below, the sum in Eq.~\eqref{eq:epr} is simplified since not all states are connected. In particular there are three contributions: from the transitions between states with $n_G=0$, from the transitions between states with $n_G=1$, and from the transitions between $n_G=0$ and 1. This gives
\begin{align}\label{eq:macEPR}
    \begin{split}
        \dot{s}_{\mathrm{mes}} = \sum_{i=1}^\infty \Bigg\{b_{i} \ln\left( \frac{\rho_{\mathrm{on}}}{\rho_{\mathrm{off}}} \right) + a_{i} \ln\left( \frac{\sigma_{\mathrm{on}}}{\sigma_{\mathrm{off}}} \right) \Bigg\} \\
        + b_1 \ln\left( \frac{\sigma_{\mathrm{on}}}{\sigma_{\mathrm{off}}} \right),
    \end{split}
\end{align}
in which $a_i = b_{i+1}-b_i$. The final term outside of the sum accounts for the flux between $(0,0)$ and $(1,0)$. This expression simplifies upon utilizing the relation $\sum_{i=1}^\infty a_i = -b_1$ and realizing that one can use $G_0(z)$ and $G_1(z)$ (from the previous section) to calculate $\sum_i b_i$ to give,
\begin{align}\label{eq:entropy-prod-rate}
\begin{split}
    \dot{s}_{\mathrm{mes}}=& \frac{(\rho_{\mathrm{on}}-\rho_{\mathrm{off}})\sigma_{\mathrm{on}} \sigma_{\mathrm{off}}}{(\sigma_{\mathrm{on}}+\sigma_{\mathrm{off}})(d+\sigma_{\mathrm{on}}+\sigma_{\mathrm{off}})}\ln\left( \frac{\rho_{\mathrm{on}}}{\rho_{\mathrm{off}}} \right),
\end{split}
\end{align}
or utilizing the definition of the mean burst size $B=\rho_{\mathrm{on}}/\sigma_{\mathrm{off}}$,
\begin{align}\label{eq:entropy-prod-rate2}
\begin{split}
    \dot{s}_{\mathrm{mes}}=& \frac{(B\sigma_{\mathrm{off}}-\rho_{\mathrm{off}})\sigma_{\mathrm{on}} \sigma_{\mathrm{off}}}{(\sigma_{\mathrm{on}}+\sigma_{\mathrm{off}})(d+\sigma_{\mathrm{on}}+\sigma_{\mathrm{off}})}\ln\left( \frac{B\sigma_{\mathrm{off}}}{\rho_{\mathrm{off}}} \right).
\end{split}
\end{align}
For full details of these calculations please see Section \ref{sec:EPRcalc}. When $\rho_{\mathrm{off}}=\rho_{\mathrm{on}}$ one finds detailed balance is satisfied with $\dot{s}_{\mathrm{mes}} = 0$, which is intuitive because the two-state gene reduces to constitutive gene expression in this case.

Eq.~\eqref{eq:entropy-prod-rate} elucidates that if either $\sigma_{\mathrm{on}}$ or $\sigma_{\mathrm{off}}$ are zero then the steady state is also in detailed balance since one is coupled to only a single mRNA reservoir. This formula also tells us that as $\rho_{\mathrm{off}}\to 0$ the entropy production becomes infinite, and the process becomes completely irreversible---albeit logarithmically slowly. Therefore, the often employed assumption of $\rho_{\mathrm{off}}= 0$ actually represents the most non-equilibrium form of the two-state model \cite{larsson2019genomic,suter2011mammalian}, as is clear from Crooks' fluctuation theorem \cite{crooks1999entropy}. Curiously, when $\rho_{\mathrm{off}}=0$, \textit{even though the dynamics are completely irreversible, the marginal mRNA dynamics still satisfy detailed balance at steady state}. This fact should provide a motivation for the study of coarse graining in mechanistic mRNA expression models that have an effective off state without being thermodynamically impossible---as coarse graining via lumping should give effective models that provide a lower bound on the EPR, not an infinite value of the EPR \cite{bo2014entropy}.

\subsection*{Entropy production of transcription in real genes}

\noindent To investigate the role that $\dot s_{\mathrm{mes}}$ may play as a selection criterion in the kinetic parameters of transcription, we took seven datasets of kinetic parameters from the recent literature and calculated $\dot s_{\mathrm{mes}}$ across thousands of genes. These datasets consisted of:
\begin{itemize}
    \item Two datasets (both of around 1,500 genes) split into G1 and G2M cell-cycle stages in mouse fibroblasts (from \cite{sukys2025cell}, based on data in \cite{riba2022cell}).
    \item One recent dataset (of around 5,600 genes) in mouse fibroblasts \cite{ramskold2024single}.
    \item Four datasets (each of around 5,000-6,000 genes) across both mouse fibroblasts and mouse embryonic stem cells \cite{larsson2019genomic}.
\end{itemize}
Only the data sets from \cite{sukys2025cell}: (i) split the data into cell-cycle stages (removing some of the cell-to-cell variability); and (ii) excluded genes from the analysis that were not best modeled as a telegraph model. This leads to some discrepancies between the kinetic parameter distributions reported in \cite{sukys2025cell} and \cite{larsson2019genomic,ramskold2024single}. Notably, in the data of \cite{larsson2019genomic,ramskold2024single} there is a hard inferential lower bound imposed on the burst size at $B=1$, which may not be a good boundary choice for many genes. Here, we give more credence to results based on \cite{sukys2025cell} due to the extra steps taken to ensure quality of the inferred parameters, but we still report results from the other datasets in the SI.

\begin{figure}[ht]
    \centering
    \includegraphics[width=.5\textwidth]{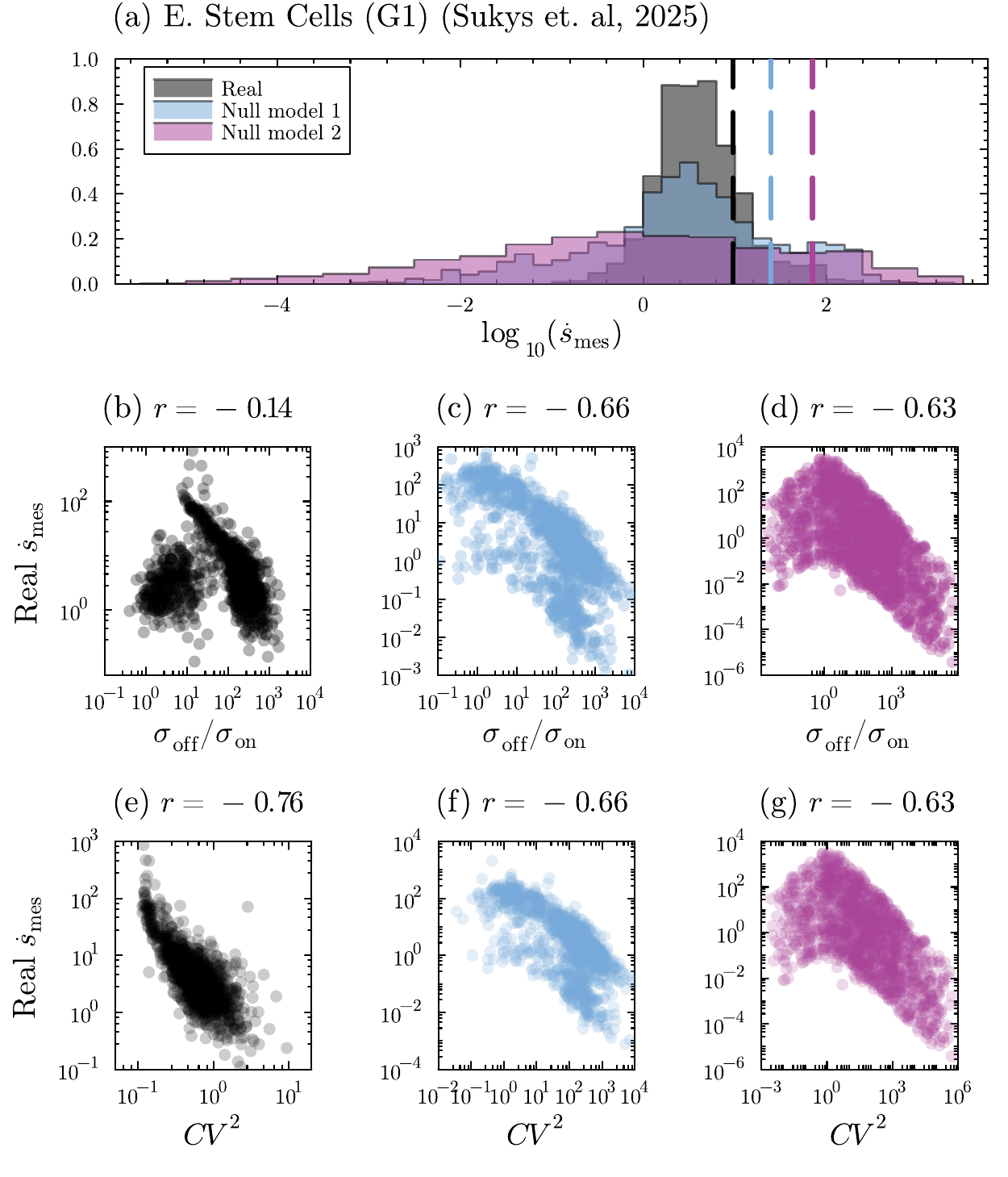}
    \caption{\textbf{Investigating detailed balance breaking via $\dot s_{\mathrm{mes}}$ between real parameters sets and null models 1 and 2 for G1 cell-cycle stage mouse fibroblasts in \cite{sukys2025cell}.} (a) Histogram over $\dot s_{\mathrm{mes}}$ for 1,436 real parameters sets (in gray), 1,436 parameter sets under null model 1 (in blue), and 1,436 parameter sets under null model 2 (in pink). Dashed lines show the respective mean values for each histogram. (b) Plot of $\dot s_{\mathrm{mes}}$ versus $\sigma_{\mathrm{off}}/\sigma_{\mathrm{on}}$ for real parameter sets. Each marker represents the position of a single gene in the phase space, and the $r$ value in the plot title is Pearson's correlation coefficient between the logarithm of the $x$ and $y$ variables. (c)-(d) Plots of $\dot s_{\mathrm{mes}}$ versus $\sigma_{\mathrm{off}}/\sigma_{\mathrm{on}}$ under null models 1 and 2 respectively. (e) Plot of $\dot s_{\mathrm{mes}}$ versus $CV^2$ for real parameter sets. (f)-(g) Plots of $\dot s_{\mathrm{mes}}$ versus $CV^2$ under null models 1 and 2 respectively.}
    \label{fig:fig3}
\end{figure}

All of these datasets report values of $\{\rho_{\mathrm{on}}, \sigma_{\mathrm{off}},\sigma_{\mathrm{on}} \}$. To obtain values for the mRNA degradation rate, $d$, we cross-referenced gene names with the mouse mRNA degradation rate dataset from \cite{sharova2009database}. Note that these degradation rates were measured in mouse embryonic stem cells, and here we assume that they are also a good approximation of the degradation rates of mRNA in mouse fibroblast cells. Finally, since values of $\rho_{\mathrm{off}}$ have not been inferred (due to parameter indistinguishability), we set $\rho_{\mathrm{off}}=10^{-3}s^{-1}$. Note that this value of $\rho_{\mathrm{off}}$ is quite arbitrary as long as it is chosen such that $\rho_{\mathrm{off}}\ll \rho_{\mathrm{on}}$ for two reasons. First, the logarithmic dependence of $\dot s_{\mathrm{mes}}$ on $\rho_{\mathrm{off}}$ implies that making $\rho_{\mathrm{off}}$ orders of magnitude smaller only changes $\dot s_{\mathrm{mes}}$ by a multiplicative factor of order $\mathcal{O}(1)$. In our calculation of the lower bound on the EPR of mRNA transcription below, this means that $\dot s_{\mathrm{mes}}$ is relatively insensitive the value of $\rho_{\mathrm{off}}$, so long as it is chosen such that $\rho_{\mathrm{off}}\ll \rho_{\mathrm{on}}$. Second, in our subsequent analyses we either: (i) compare the patterns seen in the real parameter sets to randomly drawn parameter sets, but for the same value of $\rho_{\mathrm{off}}$ (in Fig.~\ref{fig:fig3}); or (ii) compare values of $\dot s_{\mathrm{mes}}$ across parameter sets where the interest is in the differences in $\dot s_{\mathrm{mes}}$ across parameter space.

\subsection*{Statistical Null Models}
\noindent For comparison to the real kinetic parameters sets introduced above, we introduce two different null models that serve as ``neutral'' comparisons when making claims about how physically realized parameter sets may impact the EPR.

\textit{Null model 1:} To assess the degree to which specific combinations of real parameters affect the EPR we introduce a null model in which parameters are independently randomly sampled from the experimental values (with replacement) for each gene in the dataset. This ensemble is then comprised of the same number of ``genes'' but where all interdependencies between kinetic parameters have been removed. Therefore, comparisons between the real data and null model 1 test whether specific combinations of parameters have an influence on the magnitude of $\dot s_{\mathrm{mes}}$, isolating the influence of the \textit{combinations of physically realized parameters} on the EPR.

\textit{Null model 2:} To assess the degree to which specific kinetic parameters affect the EPR we introduce a null model in which the kinetic parameters themselves are \textit{randomly sampled} from neutral probability distributions (i.e., not from the data). Because each kinetic parameter varies over several orders of magnitude, we sampled each parameter from a log-uniform distribution where the interval of the distribution is defined by the minimum and maximum values of each kinetic parameter in the data. This ensemble is also comprised of the same number of ``genes'' but where the kinetic parameters are randomly sampled. Therefore, comparisons between the real data and null model 2 test whether the specific values of kinetic parameters have significant influence over $\dot s_{\mathrm{mes}}$, or whether Eq.~\eqref{eq:entropy-prod-rate2} and similar order of magnitude values of the kinetic parameters are enough to explain the magnitude of $\dot s_{\mathrm{mes}}$. That is, null model 2 isolates the impact of the \textit{distribution of kinetic parameters} on the EPR.

\subsection*{Relationship between $\dot{s}_{\mathrm{mes}}$ and real parameters}
\noindent In Fig.~\ref{fig:fig3}, we compare $\dot s_{\mathrm{mes}}$ calculated from real parameter sets for mouse fibroblast cells in the G1 cell-cycle stage \cite{sukys2025cell} to $\dot s_{\mathrm{mes}}$ evaluated under the null models. We find several key features, that are broadly exhibited across all seven datasets in Figs.~\ref{fig:sukysg2m}-\ref{fig:2019fibcast}. First, as shown in Fig.~\ref{fig:fig3}(a), the real parameter sets have EPRs that cluster much more tightly than the null models, meaning that real gene expression is characterized by a more tightly regulated EPR. Second, there is a systematically lower value mean of $\dot s_{\mathrm{mes}}$ for the real parameter sets than the null models. This indicates: (i) that transcription has kinetics that have lower $\dot s_{\mathrm{mes}}$ than occurs through randomly assigned but real kinetic values (null model 1) and (ii) that the real values of kinetic parameters reduce the EPR compared to kinetic parameters that are fully randomized (null model 2). Third, we looked at $\dot s_{\mathrm{mes}}$ against $\sigma_{\mathrm{off}}/\sigma_{\mathrm{on}}$. Comparing Figs.~\ref{fig:fig3}(b)-(d), we find that the real parameter sets lead to two distinct clusters with an overall small correlation coefficient (although the right-hand cluster has a clearly has a negative correlation), while the null models exhibit a stronger negative correlation without the clear demarcation of two distinct clusters. For real parameter sets, these results highlight that there are two distinct types transcriptional behavior, and in the burstier of these groups that increased burstiness corresponds to decreased $\dot s_{\mathrm{mes}}$. Finally, we use the formula for the coefficient of variation squared in Eq.~\eqref{eq:cv2_full} to investigate the relationship between $\dot s_{\mathrm{mes}}$ and mRNA noise. Comparing Figs.~\ref{fig:fig3}(e)-(g), we show that real parameter sets correspond to an EPR that is correlates negatively with mRNA expression noise. This correlation is weaker, albeit still present, for randomized parameter sets. In general, burstier behaviors correspond to reduced values of $\dot s_{\mathrm{mes}}$.

\begin{figure}[ht]
    \centering
    \includegraphics[width=.48\textwidth]{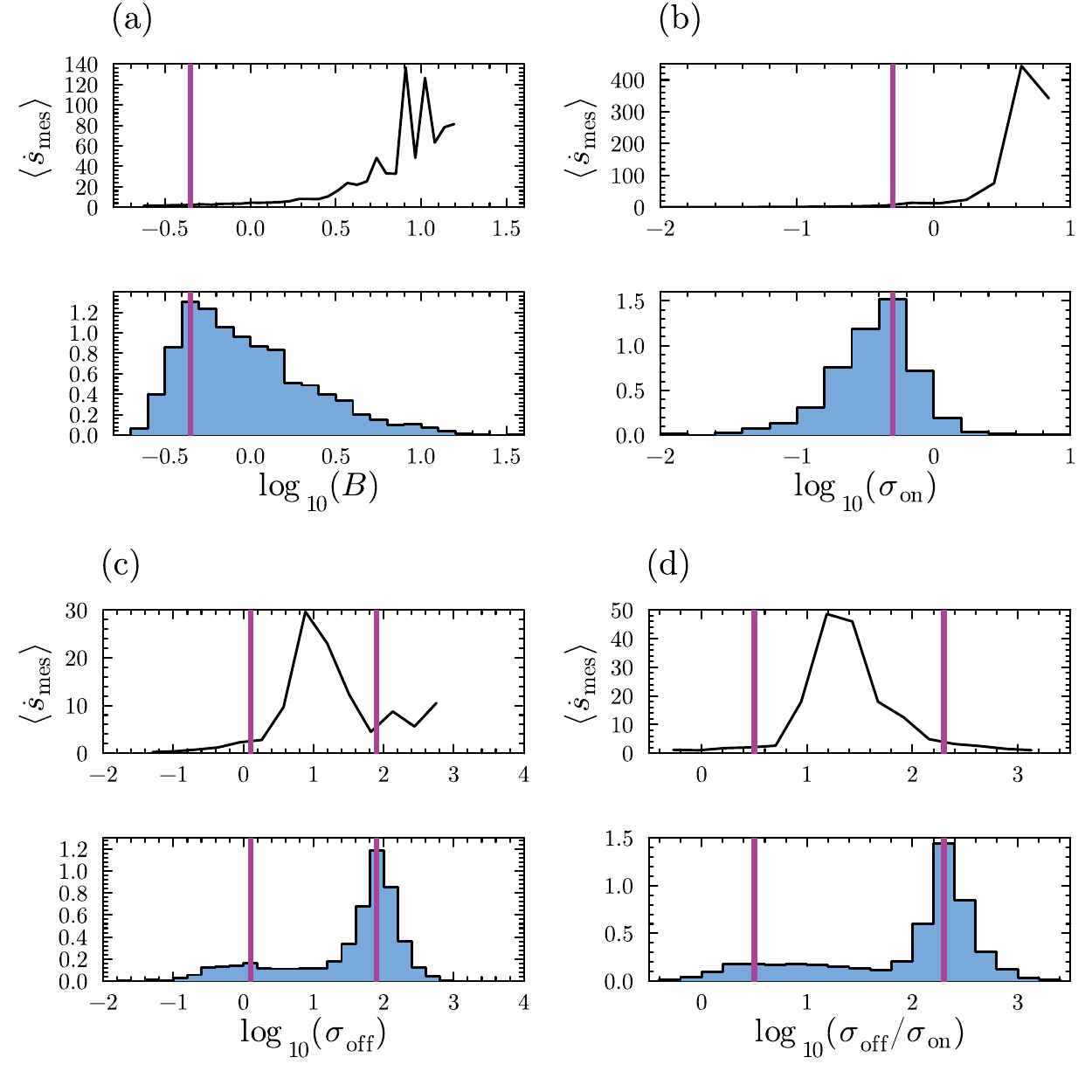}
    \caption{\textbf{Assessing the dependence of $\langle\dot s_{\mathrm{mes}}\rangle$ against real kinetic parameter combinations across 1,436 genes for G1 cell-cycle stage mouse fibroblasts in \cite{sukys2025cell}.} Here, $\langle \dot s_{\mathrm{mes}}\rangle$ is the mean EPR for binned values of \textit{real kinetic parameters sets}. In all plots, pink lines correspond to the mode of the frequency distribution over a given kinetic parameter. (a) Binning genes across $\log_{10}(B)$ shows an increasing dependence of $\langle \dot s_{\mathrm{mes}}\rangle$ on $\log_{10}(B)$. The modal value of $\log_{10}(B)$ occurs well before the rapid increase of $\langle \dot s_{\mathrm{mes}}\rangle$. (b) Binning genes across $\log_{10}(\sigma_{\mathrm{on}})$, with similar conclusions as for plot (a). (c) Binning genes across $\log_{10}(\sigma_{\mathrm{off}})$ shows a surprising feature---that the modal values of $\log_{10}(\sigma_{\mathrm{off}})$ occur before and after the peak in $\langle \dot s_{\mathrm{mes}}\rangle$. (d) For $\log_{10}(\sigma_{\mathrm{off}}/\sigma_{\mathrm{on}})$ similar conclusions are reached as for panel (c).}
    \label{fig:fig5}
\end{figure}

To investigate the average dependence of $\dot s_{\mathrm{mes}}$ for different values of kinetic parameters, we explore the dependence of $\dot s_{\mathrm{mes}}$ binned across kinetic parameters. To perform this binning procedure, we take one of the kinetic parameters, say $B$, split it into $M$ equally spaced bins (in log-space) and then average the corresponding values of $\dot s_{\mathrm{mes}}$ in each bin to give,
\begin{align}
    \langle \dot s_{\mathrm{mes}}(B_i)\rangle = \frac{1}{N_i}\sum_{j=1}^{N_i} \dot s_{\mathrm{mes},j},
\end{align}
in which $i$ denotes the bin index, $B_i$ is the midpoint value of $B$ in bin $i$, $N_i$ is the number of values of $B$ in bin $i$, and $\dot s_{\mathrm{mes},j}$ is the $j^{\mathrm{th}}$ value of $\dot s_{\mathrm{mes}}$ in bin $i$. The number of bins, $M$, varies depending on the distribution of the distribution of a given kinetic parameter. For a broadly and relatively evenly distributed kinetic parameter, a larger value of $M$ was selected (e.g., $M=40$ in Fig.~\ref{fig:fig5}(a)). For a sharply peaked kinetic parameter, a smaller value of $M$ was selected, such that regions with low density may still have a few values to average over (e.g., $M=15$ in Fig.~\ref{fig:fig5}(c)).

\begin{figure}[ht]
    \centering
    \includegraphics[width=.48\textwidth]{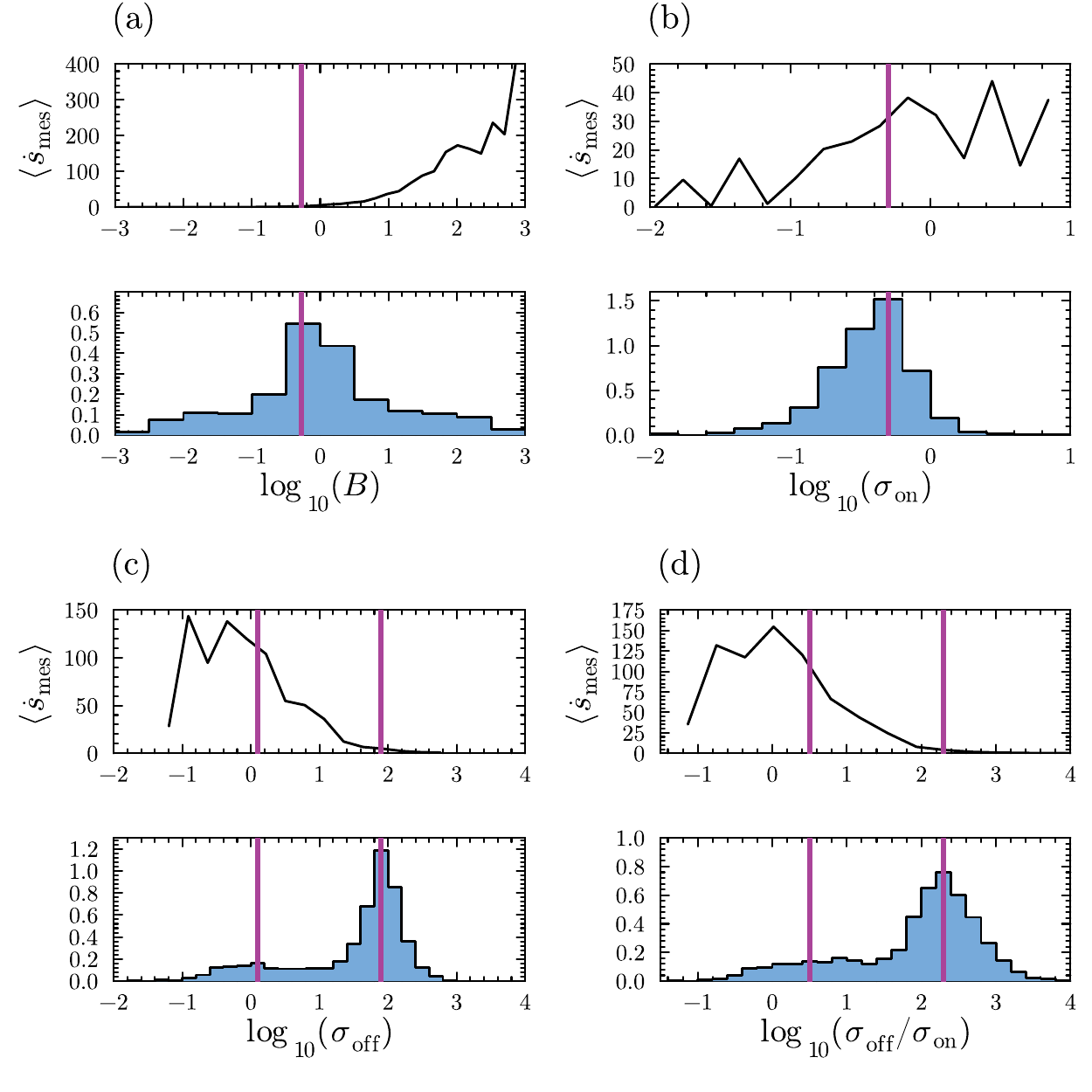}
    \caption{\textbf{Assessing the dependence of $\langle\dot s_{\mathrm{mes}}\rangle$ against realizations of \textit{null model 1} for G1 cell-cycle stage mouse fibroblasts in \cite{sukys2025cell}.} (a) Binning genes across $\log_{10}(B)$ shows little difference to Fig.~\ref{fig:fig4}(a). (b) Binning across $\log_{10}(\sigma_{\mathrm{on}})$ now results in the modal value of $\sigma_{\mathrm{on}}$ being located at higher $\langle\dot s_{\mathrm{mes}}\rangle$ than for the real kinetic parameter combinations. (c) Binning across $\log_{10}(\sigma_{\mathrm{off}})$ shows one mode coinciding with a peak in the $\langle\dot s_{\mathrm{mes}}\rangle$. (d) For $\log_{10}(\sigma_{\mathrm{off}}/\sigma_{\mathrm{on}})$ similar conclusions are reached as for panel (c).}
    \label{fig:fig-null-bin1}
\end{figure}

\begin{figure}[ht]
    \centering
    \includegraphics[width=.48\textwidth]{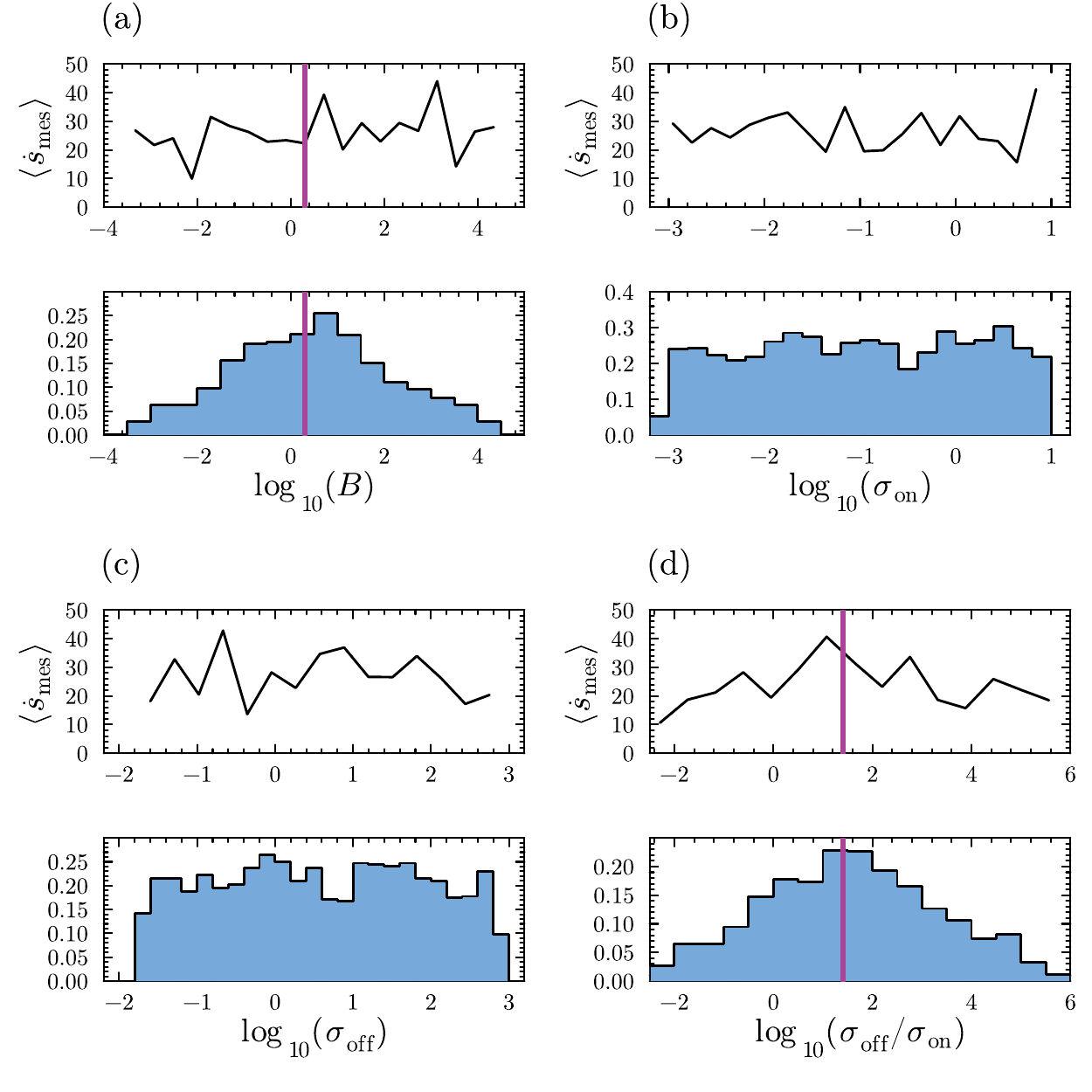}
    \caption{\textbf{Assessing the dependence of $\langle\dot s_{\mathrm{mes}}\rangle$ against realizations of \textit{null model 2} for G1 cell-cycle stage mouse fibroblasts in \cite{sukys2025cell}.} Across (a)-(d) we see little relationship between $\langle\dot s_{\mathrm{mes}}\rangle$ and kinetic parameters.}
    \label{fig:fig-null-bin2}
\end{figure}

In Fig.~\ref{fig:fig5}, we show the binning-based analysis on G1 cell-cycle stage mouse fibroblasts. Each panel shows (above) the binned $\langle \dot s_{\mathrm{mes}}\rangle$ values across the measured range of kinetic parameters, and (below) the frequency distribution of parameter values. This allows for a side-by-side comparison to observe whether there is a relationship between $\langle \dot s_{\mathrm{mes}}\rangle$ and the distribution of parameter values. In Figs.~\ref{fig:fig-null-bin1} and \ref{fig:fig-null-bin2} we replicate the same figure but for null models 1 and 2 respectively.

There are several key findings for across Figs.~\ref{fig:fig5}-\ref{fig:fig-null-bin2} centered around a common theme: \textit{that peaks in the frequency distributions of kinetic parameters describe areas of low $\langle \dot s_{\mathrm{mes}}\rangle$}. This hints that low values of $\langle \dot s_{\mathrm{mes}}\rangle$ are characteristic of real transcription. This is most strikingly seen in Figs.~\ref{fig:fig5}(c) and (d), where the peaks of the frequency distribution are found on either side of a peak in $\langle \dot s_{\mathrm{mes}}\rangle$. This same feature is also observed for G2M cell-cycle stage mouse fibroblasts in Fig.~\ref{fig:bmsg2}(c). To assess whether the choice of averaging of $\dot s_{\mathrm{mes}}$ affects this observed phenomenology, we explored using the median, minimum and maximum of $\dot s_{\mathrm{mes}}$ in a given bin. In Fig.~\ref{fig:medminmax}, we show that using these other metrics recapitulates the same phenomenon, and the peak in a given metric does not coincide with a peak in the frequency distribution. 

For the null models, areas of low $\langle \dot s_{\mathrm{mes}}\rangle$ are not necessarily indicative of peaks in kinetic parameter values. For null model 1, we see in Fig.~\ref{fig:fig-null-bin1}(b) that the modal value of $\sigma_{\mathrm{on}}$ no longer precedes the increase in $\langle\dot s_{\mathrm{mes}}\rangle$. Figs.~\ref{fig:fig-null-bin1}(c)-(d) show that although the more pronounced of the two modes is located in an area of relatively low $\langle\dot s_{\mathrm{mes}}\rangle$ the wider left-hand mode is located close to the peak in $\langle\dot s_{\mathrm{mes}}\rangle$---with a value of $\langle\dot s_{\mathrm{mes}}\rangle$ approximately 100 times higher than for the real parameter sets. This indicates that in the burstier regime of $\sigma_{\mathrm{off}}\gg\sigma_{\mathrm{on}}$ that $\langle\dot s_{\mathrm{mes}}\rangle$ will be low regardless of the values of $B$ and $d$, whereas genes for which $\sigma_{\mathrm{off}}\approx \sigma_{\mathrm{on}}$ obtain low $\langle\dot s_{\mathrm{mes}}\rangle$ by varying the entire combination of parameter values, including $B$ and $d$. Note that the values of $\langle\dot s_{\mathrm{mes}}\rangle$ are in general higher for null model 1 than the real kinetic parameter sets as shown in Fig.~\ref{fig:fig3}(a). For null model 2, we see that fully randomized parameters remove the dependence of $\langle\dot s_{\mathrm{mes}}\rangle$ on the kinetic parameters. We observe the same properties of the null models for G2M cell-cycle stage mouse fibroblast genes in Figs.~\ref{fig:g2mnull1} and \ref{fig:g2mnull2}.

These findings point towards large values of $\dot s_{\mathrm{mes}}$ being disfavored for a genes transcriptional kinetics. One could think of this as a mesoscopic analog of \textit{energy expenditure minimization} \cite{govern2014optimal} or the \textit{principle of least dissipation} \cite{han2008least}, both of which at a microscopic level state that to be energy efficient is more beneficial. For mesoscopic systems, such principles have been hinted to in recent work \cite{ray2025large}. So far, $\dot s_{\mathrm{mes}}$ has been interpreted as a measure of irreversibility, and not as an entropy production rate in the thermodynamic sense. In the next section we assess whether these results have any thermodynamic relevance.
 
For datasets from \cite{larsson2019genomic} and \cite{ramskold2024single} (see Figs.~\ref{fig:bmsr}-\ref{fig:bms2019fcast}) the range of values of $\langle \dot s_{\mathrm{mes}}\rangle$ found across each kinetic parameter was found to be far less that the dataset in \cite{sukys2025cell}, meaning that the phenomenon of high parameter abundance implying low $\langle \dot s_{\mathrm{mes}}\rangle$ is not an obvious conclusion. We suspect that the imposition of the inferential lower bound $B=1$ has some influence on the differing conclusions between the data from \cite{sukys2025cell} and those from \cite{larsson2019genomic,ramskold2024single}.

\subsection*{Thermodynamic bounds from mesoscopic EPR}

\noindent There is a large literature that discusses how coarse-grained systems provide lower bounds on the EPR of the corresponding microscopic system \cite{egolf2000equilibrium,blythe2008reversibility,gomez2008lower,england2013statistical,bo2017multiple,jia2016model,skinner2021estimating}. Using these references, in combination with previous work on housekeeping free energy usage \cite{ge2013dissipation}, it follows that
\begin{align}\label{eq:Fdis}
    \overline{f_{\mathrm{dis}}} \geq T \dot{s}_{\mathrm{mes}},
\end{align}
where $\overline{f_{\mathrm{dis}}}$ is the average rate of free energy dissipation per cell necessary to keep the system out of equilibrium, and $T$ is the temperature of the cell. Here, if we assume that the two-state gene model is a coarse-grained model of all the microscopic processes making up the transcription of a single gene, then Eqs.~\eqref{eq:entropy-prod-rate} and \eqref{eq:entropy-prod-rate2} should provide a lower bound on the free energy rate necessary to keep a gene transcribing with parameters $\{\rho_{\mathrm{on}},\rho_{\mathrm{off}}, \sigma_{\mathrm{on}}, \sigma_{\mathrm{off}}, d\}$. Based on England's calculations in \cite{england2013statistical}, given that a coarse-grained birth-death process seems provides a relatively tight lower bound on the thermodynamics of a proliferating population of cells, we may be cautiously optimistic that the two-stage model of gene expression can provide a lower energy bound for the transcription of a single gene.

To assess the tightness of $T \dot{s}_{\mathrm{mes}}$ as a lower bound on $\overline{f_{\mathrm{dis}}}$, we can make an order of magnitude estimate on the energy needed for transcription based on measured values. Let us say that the energy needed for transcription is in the turnover of producing mRNA transcripts from pre-existing nucleotides. The energy cost of this is 2 phosphorylation events per chain elongation step (i.e., per nucleotide) \cite{lynch2015bioenergetic}, and the energy of a single phosphorylation event is around $10 k T$ \cite[Sec.~\textit{What is the energy in transfer of a phosphate group?}]{milo2015cell}. Therefore, for a gene of length $L$, with kinetic parameters $\{\rho_{\mathrm{on}},\rho_{\mathrm{off}}, \sigma_{\mathrm{on}}, \sigma_{\mathrm{off}}\}$ (where $\rho_{\mathrm{off}}\ll \rho_{\mathrm{on}}$) the energy usage per second is approximately
\begin{align}
    \tilde f \approx 10 k T\times L\times \frac{\rho_{\mathrm{on}} \sigma_{\mathrm{on}}}{\sigma_{\mathrm{on}}+\sigma_{\mathrm{off}}}.
\end{align}
For a gene of length $1k\cdot bp$, and mean values of $\sigma_{\mathrm{on}} = 0.47 \,s^{-1}, \sigma_{\mathrm{off}} = 77.25 \,s^{-1}, \rho_{\mathrm{on}} = 62.85 \,s^{-1}$ (from G1 cell-cycle stage mouse fibroblast cells) this results in $\tilde f\approx 3800 k T$. Given that the largest values of $\dot{s}_{\mathrm{mes}}$ for the same dataset are of order $\mathcal{O}(10^3)$ with a mean of order $\mathcal{O}(10^1)$ (see Fig.~\ref{fig:fig3}), this shows that one cannot interpret $\dot{s}_{\mathrm{mes}}$ as having thermodynamic relevance. Recent work in \cite{kolchinsky2024thermodynamic} has shown that the results of England in \cite{england2013statistical} should be exercised with caution. The results of this section are in line with these findings.

\begin{figure}[ht]
    \raggedright
    \includegraphics[width=.49\textwidth]{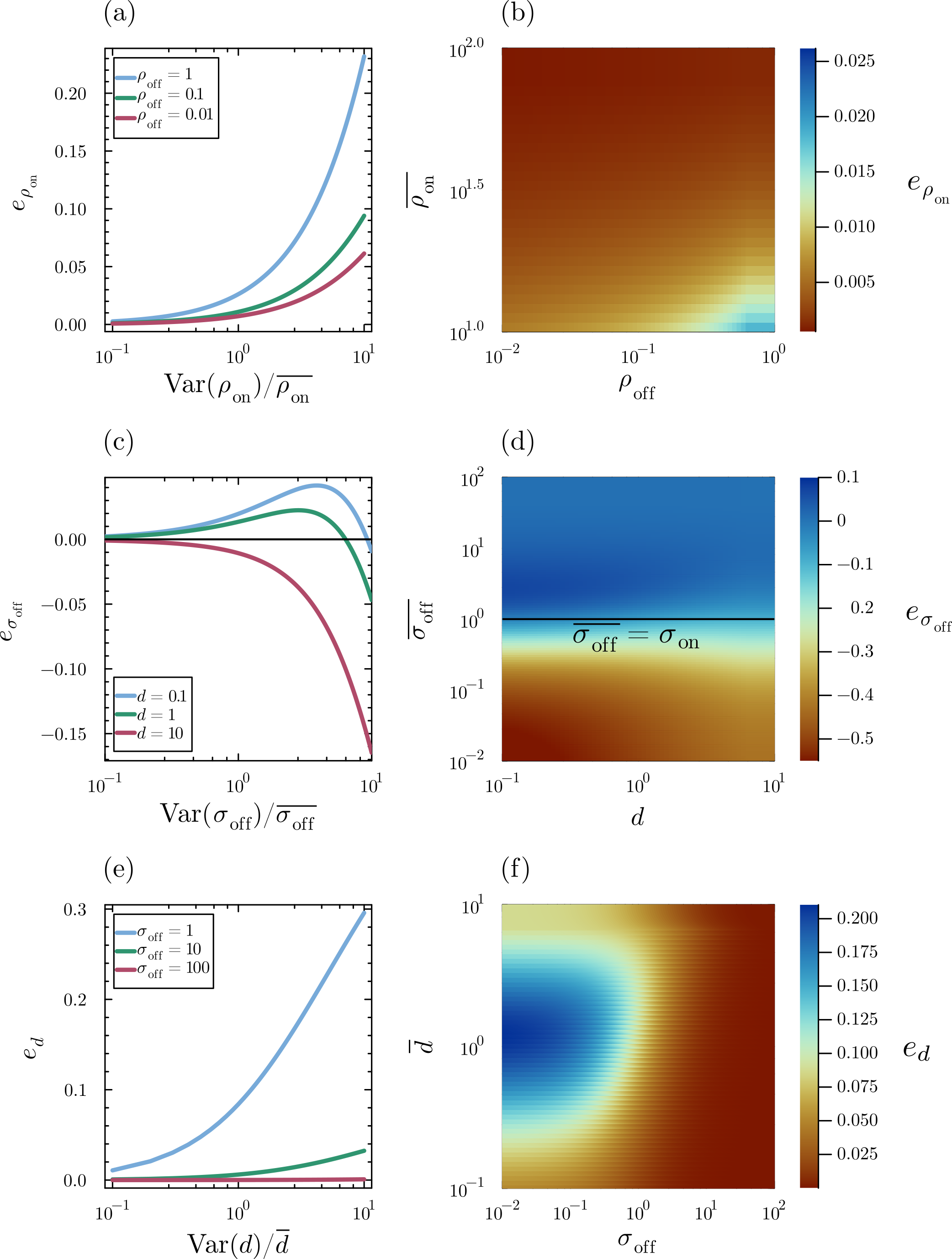}
    \caption{\textbf{Deviations induced in $\dot{s}_{\mathrm{mes}}$ due to cell-to-cell variability.} (a), (c) and (e) show how the relative error changes for kinetic rates $\rho_{\mathrm{on}},\sigma_{\mathrm{off}}$ and $d$ for constant means and increasing variance for slices of parameter space. (b), (d) and (f) show density plots of the error across all relevant dimensions of parameter space, for fixed $\overline\theta = \mathrm{Var}(\theta)$. In (c) $\sigma_{\mathrm{on}} = 3$, in (d) $\sigma_{\mathrm{on}}=1$, in (e) $\sigma_{\mathrm{on}}=1$ and in (f) $\sigma_{\mathrm{on}}=1$.}
    \label{fig:fig_ext}
\end{figure}

\subsection*{Entropy production of transcription in a population of cells}\label{sec:EPRpop}

\noindent In the previous section, we focused on the EPR of transcription of a single gene assuming a canonical two-state gene. Scaling-up this calculation up for a population of identical cells is straightforward---the EPR is simply multiplied by the number of cells $N$ (if the cells are assumed to be non-interacting). However, populations of cells vary in many features, such as cell size, pH and morphology. Therefore, inferring a single set of kinetic parameters for a population with variable kinetics can lead to significant inferential error \cite{lim2015quantitative,grima2023quantifying,gorin2024new} (see Fig.~\ref{fig:schematic}(b)). By extension, the subtleties of extrinsic noise may lead to properties of reversibility of an ensemble that are non-trivially related to those of a single cell.

Following \cite{ham2020extrinsic,grima2023quantifying}, we assume that the kinetic rates of the two-state model are gamma distributed---since the mean and variance of the distribution can be tuned independently and the gamma distribution is defined for positive real numbers (kinetic rates are, by definition, positive). Then, each cell in the population has kinetic parameters which are drawn from these distributions. We investigate the effect of having this extrinsic noise parameter-by-parameter, for two reasons: (i) it allows for the isolation of the effects of noise on that parameter on the EPR and (ii) it allows for analytic results for $\dot s_{\mathrm{mes}}$ across a variable population. Below, $\theta$ represents one of $\sigma_{\mathrm{on}},\sigma_{\mathrm{off}},\rho_{\mathrm{on}},d$ and $\rho_{\mathrm{off}}$ in a single cell. The gamma distribution has the probability density function
\begin{align}\nonumber
    f(\theta;\alpha,\beta) = \frac{\beta^\alpha}{\Gamma(\alpha)}\theta^{\alpha-1}e^{-\beta \theta},
\end{align}
where $\alpha$ represents the shape parameter and $\beta$ the inverse scale parameter, in which the mean is $\overline\theta = \alpha/\beta$ and the variance as $\mathrm{Var}(\theta)=\alpha/\beta^2$. We use $\overline x$ to denote the average of $x$ over a population of cells. For comparison with the single cell case, $\dot{s}_{\mathrm{mes}}[\theta]$ is the average mesoscopic EPR \textit{per cell}, an intrinsic quantity, given by
\begin{align}\label{eq:extint}
    \dot{s}_{\mathrm{mes}}[\theta;\alpha,\beta] =\int_0^\infty \dot{s}_{\mathrm{mes}} f(\theta;\alpha,\beta)\mathrm{d}\theta,
\end{align}
where $\dot{s}_{\mathrm{mes}}$ in the integrand is taken from Eq.~\eqref{eq:entropy-prod-rate}, and the functional parameter $\theta\in\{\sigma_{\mathrm{on}},\sigma_{\mathrm{off}},\rho_{\mathrm{on}},d,\rho_{\mathrm{off}}\}$ denotes the parameter that is varying over the population. We also define
\begin{align}\label{eq:errorfn}
    e_{\theta}(\alpha,\beta) = \frac{\dot{s}_{\mathrm{mes}}[\theta;\alpha,\beta] - \dot{s}_{\mathrm{mes}}}{\dot{s}_{\mathrm{mes}}}
\end{align}
as the relative error from the case of identical cells if $\theta$ is varied as a gamma distribution in the population with parameters $\alpha,\beta$ and $\dot{s}_{\mathrm{mes}}$ is evaluated at $\overline\theta$. In principle, this process can be extended for higher-order moments of $\dot{s}_{\mathrm{mes}}$, although it is more difficult to obtain analytic expressions.

For all five kinetic parameters, the integrals in Eq.~\eqref{eq:extint} are solvable, typically using special functions.
Note that, since both $\sigma_{\mathrm{off}}$ and $\sigma_{\mathrm{on}}$ occur in $\dot{s}_{\mathrm{mes}}$ symmetrically, only the results for $\sigma_{\mathrm{off}}$ are reported. The solutions to the integrals in Eq.~\eqref{eq:extint} are rather cumbersome and so we report these expressions for all kinetic parameters in the Section \ref{sec:ext_noise_a}, and show one of the simpler results for noise on the degradation rate
\begin{align}\nonumber
    \dot{s}_{\mathrm{mes}}[d] = \frac{\exp{\left(\Sigma\beta\right)}\beta^\alpha\delta\sigma_{\mathrm{on}} \sigma_{\mathrm{off}} \Gamma(1-\alpha,\Sigma\beta)\ln\left(\frac{\rho_{\mathrm{on}}}{\rho_{\mathrm{off}}}\right)}{(\sigma_{\mathrm{on}}+\sigma_{\mathrm{off}})^{2-\alpha}}
\end{align}
which has the corresponding relative error of
\begin{align}\nonumber
    e_{d} = \exp{\left(\Sigma\beta\right)} (\alpha+\Sigma\beta)E_{\alpha}(\Sigma\beta)-1,
\end{align}
where $E_{\alpha }(\cdot)$ is the exponential integral, and the dependence on $\alpha$ and $\beta$ on $e_d$ has been omitted for brevity. The functional dependence of the error function is generally only in one or two kinetic parameters. As seen in Section \ref{sec:ext_noise_a}: $e_{\rho_{\mathrm{on}}}$ is a function of $\rho_{\mathrm{off}}$; $e_{\rho_{\mathrm{off}}}$ is a function of $\rho_{\mathrm{on}}$; $e_{\sigma_{\mathrm{off}}}$ is a function of $d$ and $\sigma_{\mathrm{on}}$; and $e_d$ is a function of $\sigma_{\mathrm{on}}$ and $\sigma_{\mathrm{off}}$. For $\rho_{\mathrm{off}}\ll \rho_{\mathrm{on}}$, the relative error $e_{\rho_{\mathrm{off}}}$ is close to zero, and hence is excluded from the more prominent analyses shown below.

The results of the extrinsic noise analysis are shown in Fig.~\ref{fig:fig_ext}. The line plots in Figs.~\ref{fig:fig_ext}(a), (c) and (e) show how the relative error changes as a function of the variance of the extrinsic noise for fixed mean, while the density plots show the errors for given parameter sets at fixed $\overline\theta = \mathrm{Var}(\theta)$. Generally, it is found that having cell-to-cell variability in the population increases $\dot s_{\mathrm{mes}}$, and that in general the system is further away from detailed balance---see Fig.~\ref{fig:fig_ext}(a), (b), (e) and (f). The exception to this rule is that when the noise is on $\sigma_{\mathrm{off}}$, $e_{\sigma_{\mathrm{off}}}$ can be positive or negative depending on the values of $\sigma_{\mathrm{on}}$ and $d$---see Fig.~\ref{fig:fig_ext}(c) and (d). The result is that the EPR of a system with extrinsically varying kinetic parameters in a population does not correspond with the ``representative cell'' comprising the mean of all kinetic parameters.

The lack of correspondence between a representative cell and population averages may seem counterintuitive---especially for a population of non-interacting cells---but $\dot{s}_{\mathrm{mes}}$ is not linear in any of the kinetic parameters, therefore in general $\dot{s}_{\mathrm{mes}}[\theta]\neq\dot{s}_{\mathrm{mes}}(\overline\theta)$. When cell-to-cell variability is present in $\rho_{\mathrm{on}}$ or $d$, it is found that extrinsic variability only increases $\dot{s}_{\mathrm{mes}}$. For $\rho_{\mathrm{on}}$, this is to a lesser extent, and in the case of Fig.~\ref{fig:fig_ext}(b), $\dot{s}_{\mathrm{mes}}$ continually increases as $\rho_{\mathrm{off}}$ gets smaller. For $d$, this is to a greater extent, with over a $20\%$ increase in $\dot{s}_{\mathrm{mes}}[d]$ in cases where $\sigma_{\mathrm{off}}$ is small and $\overline d$ is of the same order as $\sigma_{\mathrm{on}}$ (see Fig.~\ref{fig:fig_ext}(f)). Additionally, it is found that for $\dot{s}_{\mathrm{mes}}[d]$ and $\dot{s}_{\mathrm{mes}}[\rho_{\mathrm{on}}]$ that increasing the variance with the mean fixed leads to monotonically increases the relative error (see Figs.~\ref{fig:fig_ext}(a) and (c)).

Perhaps the most interesting results come from $e_{\sigma_{\mathrm{off}}}$ ($e_{\sigma_{\mathrm{on}}}$), where not only can $\sigma_{\mathrm{off}}$ ($\sigma_{\mathrm{on}}$) variability increase $\dot{s}_{\mathrm{mes}}$ for $\overline\sigma_{\mathrm{off}}\ll \sigma_{\mathrm{on}}$ ($\sigma_{\mathrm{off}}\gg \overline\sigma_{\mathrm{on}}$), but also significantly lower it for most of the space of parameters outside that regime (see Fig.~\ref{fig:fig_ext}(d)). Additionally, in Fig.~\ref{fig:fig_ext}(c) it is shown that $e_{\sigma_{\mathrm{off}}}$ ($e_{\sigma_{\mathrm{on}}}$) is non-monotonic in the variance of the extrinsic noise, and that for small degradation rates there is a finite valued maximum. The summary is that \textit{the irreversibility inferred in a single cell can be significantly perturbed by the presence of cell-to-cell variability.} This leads to the surprising conclusion that the suppression of cell-to-cell variability is may be more beneficial on some parameters than for others, that is, if the results of the previous section on the observed avoidance of high $\dot{s}_{\mathrm{mes}}$ (i.e., mesoscopic energy expenditure minimization) are to be taken seriously. However, until data is collected on the extrinsic variation of individual kinetic parameters, we cannot be sure whether a population-level EPR provides an evolutionary motivated constraint on the expression noise of individual genes.

\section*{Discussion}\label{sec:conc}
\noindent This study highlights the role of detailed balance and irreversibility in the canonical two-state model of mRNA expression. This includes: the nature of probability fluxes in the steady state (Fig.~\ref{fig:fig2}); the analytics of entropy production for a single gene; detailed data analyses of seven kinetic parameter datasets (Figs.~\ref{fig:fig3} and \ref{fig:fig5} and corresponding figures in the SI); and the counterintuitive effects of cell-to-cell variability on the entropy production rate across a population of cells (Fig.~\ref{fig:fig_ext}). Our results have highlighted a potential mesoscopic analog of the principle of energy expenditure minimization that reveals itself through the EPR of mRNA expression \cite{govern2014optimal}. This finding is aligned with a recent study claiming the non-equilibrium systems reduce their EPR at steady state for large system sizes \cite{ray2025large}. Notably, this is not a thermodynamic phenomenon---we have shown that the lower bound on the EPR predicted by our mesoscopic analysis is not tight compared to order of magnitude estimates. Finally, our results highlight the benefits of transcriptional bursting from the perspective that a burstier behaviors can lead to a lower EPR (Fig.~\ref{fig:fig5}(d)).

One of the limitations of our study is that we consider only a two-state gene model of transcription, whereas intricate experiments have shown that transcription in higher-order organisms is best modeled by more than two gene states \cite{suter2011mammalian,bothma2014dynamic, rodriguez2019intrinsic, bartman2019transcriptional, cao2020stochastic, tunnacliffe2020transcriptional, shelansky2024single}. Recent experimental data even reveals that some genes have sub-Poissonian mRNA expression in fission yeast \cite{weidemann2023minimal}, a behavior that aligns with earlier theoretical predictions \cite{mcclure1980rate, mitarai2008generation, choubey2015deciphering}. Although the analytic framework introduced in this study can likely be extended to consider multiple gene states, there are no studies that infer transcriptional kinetics across thousands of genes for models more complex than a two-state model \cite{larsson2019genomic,ramskold2024single,sukys2025cell}. Until such datasets are widely accessible, analytic results for the EPR of more complex gene state models cannot be fully utilized. Finally, our framework assumes gamma-distributed extrinsic noise, but heavy-tailed variation (e.g., log-normal) may further alter EPR bounds. Future work could explore this numerically or via approximations, building on studies like \cite{ham2020extrinsic}.

Other recent work in the literature echoes our study. Recent research by Gehri \textit{et al.}~\cite{gehri2024entropy}
also utilizes a stochastic thermodynamic approach to study the entropy production of models of transcription---in some cases models with three gene states. Therein, they promote the directed information rate as a measure of the throughput in cellular signal processing. Their hypothesis that maximizing the directed information rate is beneficial for transcription is distinct from the data-driven hypothesis of our paper, that most genes actively try not to have a large EPR for mRNA expression. Whereas the focus in our paper is on the EPR of mRNA expression given a two-state promoter, the focus of \cite{gehri2024entropy} is in the EPR resulting from gene state transitions. They hypothesize that energy dissipation limits the information throughput in transcription---i.e., that energy needs to be expended for a greater rate of information transmission. This result is not contradictory to the results presented herein---they could even be complementary, in the case that mRNA expression with an abnormally high EPR could correspond to essential genes that have a greater information throughput. 

While we focus on the minimization of the EPR of mRNA expression as a design principle for transcription, other work has focused on the minimization of the coefficient of variation of protein expression as being beneficial---in models of genetic autoregulation \cite{bokes2017gene}. The minimization of noise at the level of mRNA expression would contradict the minimization of mRNA expression EPR due to the observed negative correlation between mRNA expression noise and EPR in real data (see Fig.~\ref{fig:fig3}(d) and corresponding figures in the SI). This perhaps hints towards different principles guiding the observed expression of mRNA and proteins.

The finding that cell-to-cell variability can amplify or suppress irreversibility interacts with the seminal work of Lestas \textit{et al.}~\cite{lestas2010fundamental} in several ways. Therein, they investigated the limits of noise suppression given thermodynamic and information theoretic constraints, and derived quantitative bounds on the degree to which noise suppression is actually possible. Their findings, that the suppression of noise in one component of a system may amplify it in another, has clear parallels to the findings presented here, in that accepting a degree of cell-to-cell variability can lead to an ensemble of cells that is more reversible than an equivalent population of identical cells. Where noise is tolerated in the cellular environment is the result of a complex evolutionary process, but this process may not depend only on the properties of a single cell, but on the properties of a population of cells. Our study hints to further evolutionary constraints, that not only is the placement of noise and its suppression important in the context of intrinsic stochasticity, but in where extrinsic noise is tolerated, and in how it is suppressed \cite{weissman2025beyond,weinreich2025population}.

In future studies, it would be interesting to study the information theoretic properties of mRNA expression with more complex gene-state architectures---especially if advances in live-seq \cite{chen2022live} and merFISH \cite{zhang2021spatially} data allow for comprehensive datasets of kinetic parameters for more microscopic models of mammalian transcription. Time-resolved mRNA sequencing would also allow for studies into the breaking of detailed balance in dynamic cellular environments, with fluctuating and correlated mRNA expression, using methods introduced in \cite{battle2016broken,lynn2021broken}. Finally, an experimental understanding of how the kinetic parameters of the telegraph model fluctuate across a population of cells could give credence to some of the implications of cell-to-cell variability on reversibility found in this study, and of other studies related to the explicit modeling of extrinsic noise in addition to intrinsic stochasticity \cite{grima2023quantifying,holehouse2020steady,battich2015control,thomas2019intrinsic}. How cells decide where to tolerate cell-to-cell variability is an important open problem.

\emph{Acknowledgements:} This work was supported by a Lou Schuyler grant from the Santa Fe Institute and National Science Foundation grants DMR-191073 and 2133863. J.H.~would like to thank Jacob Calvert, Christopher Lynn, Anish Pandya, Brandon Schlomann, Harrison Hartle, Artemy Kolchinsky and Ramon Grima for discussions and constructive feedback. Special thanks go to Ramon Grima and Artemy Kolchinsky for invaluable suggestions. Finally, J.H.~would like to thank an anonymous reviewer for emphasizing the use of statistical null models that serve as a comparison to the real data used in this study.

\emph{Data availability:} The data for telegraph model parameters used in this paper can be found in \cite{sukys2025cell,ramskold2024single,larsson2019genomic}. The data for the mRNA degradation rates can be found in \cite{sharova2009database}.

\emph{Code availability:} The underlying code for this study is not publicly available but may be made available to qualified researchers on reasonable request from the corresponding author.

\emph{Author contributions:} J.H.~conceived, designed, implemented and wrote the study.

\emph{Competing interests:} The author declares no competing interests.

\bibliographystyle{naturemag.bst}
\bibliography{biblio}

\pagebreak
\clearpage
\widetext
\begin{center}
\textbf{\large Supplementary Information}
\end{center}
\setcounter{equation}{0}
\setcounter{figure}{0}
\setcounter{section}{0}
\setcounter{table}{0}
\setcounter{footnote}{0}
\setcounter{page}{1}
\makeatletter
\renewcommand{\theequation}{S\arabic{equation}}
\renewcommand{\thefigure}{S\arabic{figure}}
\renewcommand{\thesection}{S\arabic{section}}

\section{Expressions for the Fano factor and coefficient of variation in the two-state gene model}\label{sec:ffcv2}
\noindent From the generating function $G(z)$ in the main text, one can calculate the mean and the variance of mRNA expression via,
\begin{align}
    \langle n\rangle &= G'(z)|_{z\to1},\\
    \mathrm{Var}(n) &= \left[G''(z)+G'(z)-G'(z)^2\right]_{z\to1}.
\end{align}
Then, one can find the Fano factor and coefficient of variation squared as, 
\begin{align}\label{eq:ff_full}
    \mathrm{FF} =& \frac{
    d (\sigma_{\text{off}} + \sigma_{\text{on}})(\rho_{\text{off}} \sigma_{\text{off}} + \rho_{\text{on}} \sigma_{\text{on}})
    + \rho_{\text{off}}^2 \sigma_{\text{off}} \sigma_{\text{on}}
    }{
    (\sigma_{\text{off}} + \sigma_{\text{on}})(d + \sigma_{\text{off}} + \sigma_{\text{on}})(\rho_{\text{off}} \sigma_{\text{off}} + \rho_{\text{on}} \sigma_{\text{on}})
    }
    \\\nonumber&+
    \frac{
    \rho_{\text{off}} \sigma_{\text{off}} \left((\sigma_{\text{off}} + \sigma_{\text{on}})^2 - 2 \rho_{\text{on}} \sigma_{\text{on}}\right)
    + \rho_{\text{on}} \sigma_{\text{on}} \left(\sigma_{\text{off}} \rho_{\text{on}} + (\sigma_{\text{off}} + \sigma_{\text{on}})^2\right)
    }{
    (\sigma_{\text{off}} + \sigma_{\text{on}})(d + \sigma_{\text{off}} + \sigma_{\text{on}})(\rho_{\text{off}} \sigma_{\text{off}} + \rho_{\text{on}} \sigma_{\text{on}})
    },
    \\\label{eq:cv2_full}
    CV^2 =& \frac{
    d \left[
    d (\sigma_{\text{off}}+\sigma_{\text{on}})(\rho_{\text{off}} \sigma_{\text{off}} + \rho_{\text{on}} \sigma_{\text{on}})
    + \rho_{\text{off}}^2 \sigma_{\text{off}} \sigma_{\text{on}}
    \right]
    }{
    (d + \sigma_{\text{off}} + \sigma_{\text{on}})
    \left(\rho_{\text{off}} \sigma_{\text{off}} + \rho_{\text{on}} \sigma_{\text{on}}\right)^2
    }
    \\\nonumber&+
    \frac{
    d \left[
    \rho_{\text{off}} \sigma_{\text{off}} \left((\sigma_{\text{off}}+\sigma_{\text{on}})^2 - 2\rho_{\text{on}} \sigma_{\text{on}}\right)
    + \rho_{\text{on}} \sigma_{\text{on}} \left(\sigma_{\text{off}} \rho_{\text{on}} + (\sigma_{\text{off}}+\sigma_{\text{on}})^2\right)
    \right]
    }{
    (d + \sigma_{\text{off}} + \sigma_{\text{on}})
    \left(\rho_{\text{off}} \sigma_{\text{off}} + \rho_{\text{on}} \sigma_{\text{on}}\right)^2
    }.
\end{align}
In the ``bursty limit'' of $\{\rho_{\mathrm{on}},\sigma_{\mathrm{off}}\}\gg \{\sigma_{\mathrm{off}},\rho_{\mathrm{off}},d\}$, these expressions reduce to Eqs.~\eqref{eq:ff} and \eqref{eq:cv2} in the main text.

\section{Eigenvalues of the two-state model}\label{sec:TMsol}

\noindent In this section, we solve for the eigenvalues of the master equation describing the two-state gene model in Eq.~\eqref{eq:modTM}. The reaction scheme with $\rho_{\mathrm{off}} =0$ has already been solved in time in the present form \cite{iyer2009stochasticity,wu2023solving} and additionally for the case where mRNA production is bursty \cite{cao2018linear}. Here we specify the calculation for non-zero $\rho_{\mathrm{off}}$. Physically, the eigenvalues of the master equation correspond to the inverse of the fundamental timescales governing the relaxation towards the steady state. We find the eigenvalues by imposing physical constraints on the generating function, essentially that the generating function's power series is real and non-singular for $z\in[-1,1]$.

The generating function equations corresponding to Eqs.~\eqref{eq:Meqs} are,
\begin{align}
    \partial_t G_0 &= \rho_{\mathrm{off}}  (z-1)G_0 +d(1-z)\partial_zG_0 +\sigma_{\mathrm{off}}  G_1 - \sigma_{\mathrm{on}}  G_0,\\\nonumber
    \partial_t G_1 &= \rho_{\mathrm{on}}  (z-1)G_0+d(1-z)\partial_zG_1 -\sigma_{\mathrm{off}}  G_1 + \sigma_{\mathrm{on}}  G_0,
\end{align}
where the arguments $z$ and $t$ have been dropped for brevity. Defining $G=G_0+G_1$ (not to be confused with the gene state $G$), manipulating the generating function equations leads to the PDE describing the evolution of $G$,
\begin{align}\nonumber
    \partial_z^2 G&+\frac{\partial_t^2G}{d^2(z-1)^2}+\frac{2\partial^2_{z,t}G}{d(z-1)}+\left( \frac{\rho_{\mathrm{on}}-z \rho_{\mathrm{on}}+\Sigma +\rho_{\mathrm{off}}-z \rho_{\mathrm{off}} }{d (z-1)} \right)\partial_zG\\
    &+\left(\frac{\rho_{\mathrm{on}}-z \rho_{\mathrm{on}}-d+\Sigma +\rho_{\mathrm{off}}-z \rho
   _{\mathrm{off}}}{d^2 (z-1)^2}\right)\partial_tG +\left( \frac{(z-1) \rho_{\mathrm{on}} \rho
   _{\mathrm{off}}-\rho_{\mathrm{off}} \sigma_{\mathrm{off}}-\rho_{\mathrm{on}} \sigma_{\mathrm{on}}}{d^2 (z-1)} \right) G=0.
\end{align}
Using the separation of variables ansatz $G(x,t)\sim e^{-\lambda_m t}f_m(x)$, which arises naturally from the linear structure of the master equation, leads to \cite{van1992stochastic,tauber2014critical},
\begin{align}
    x\partial_x^2f_m(x)+\left(\Sigma_m-  \frac{\rho_{\mathrm{on}} +\rho_{\mathrm{off}} }{d}x\right)\partial_xf_m(x)+\left(\frac{\rho_{\mathrm{on}} \rho_{\mathrm{off}}  x}{d^2}+\frac{a_m}{x}+b_m\right)f_m(x)=0,
\end{align}
in which we have defined $x=\delta(z-1)/d$, $\Sigma_m=\delta(\Sigma-2\lambda_m)/d^2$, $a_m= \delta^2\lambda_m(\lambda_m+d-\Sigma)/d^4$, and $b_m=\delta(\lambda_m(\rho_{\mathrm{on}} +\rho_{\mathrm{off}} )-\rho_{\mathrm{on}} \sigma_{\mathrm{on}} -\rho_{\mathrm{off}} \sigma_{\mathrm{off}} )/d^3$, and remind the reader the definitions $\Sigma=\sigma_{\mathrm{off}} +\sigma_{\mathrm{on}} $ and $\delta=\rho_{\mathrm{on}} -\rho_{\mathrm{off}} $. This ODE has two singularities, a regular singularity at $x=0$ and an irregular singularity at $x=\infty$. This means that it can be solved by the confluent hypergeometric function. One finds by appropriate transformations of the $x$ and $f_m(x)$ that the solution is given by a sum of two orthogonal confluent hypergeometric functions,
\begin{align}\nonumber
    f_m(z) =& \exp\left( \frac{\rho_{\mathrm{off}} (z-1)}{d} \right) \Big\{C_m^1(z-1)^{\lambda_m/d} {}_1F_1\left(\frac{\sigma_{\mathrm{on}} }{d},\frac{\Sigma}{d};\frac{\delta(z-1)}{d}\right)\\ &+C_m^2(z-1)^{1+\frac{\lambda_m-\Sigma}{d}}{}_1F_1\left(1-\frac{\sigma_{\mathrm{off}} }{d},2-\frac{\Sigma}{d};\frac{\delta(z-1)}{d}\right)\Big\}.
\end{align}
The second solution here is only linearly independent if the exponent is not a integer less than or equal to 0. Now, the condition on $f_m(z)$ here necessary to determine the $\lambda_m$ is that the powers of $(z-1)$ pre-multiplying the confluent hypergeometric functions must be integer powers. This is standard practice in eigenfunction solutions which are hypergeometrics, and it means that each $f_m(z)$ is real and non-singular for $z\in[-1,1]$, and corresponds to the infinite state limit of other similar solutions found in the literature for Moran-like processes \cite{ewens2004mathematical,holehouse2022exact,mckane2000mean}. Additionally, the exponential pre-factor and the hypergeometric functions themselves are already real and non-singular for finite $z$. Enforcing that $\lambda_m=d m$ and $\lambda_m-\Sigma=dm$ for integer $m\in\{1,2,3,\ldots\}$ then gives two respective sets of eigenvalues, $\lambda_m^1 = d m$ and $\lambda_m^2 = dm+\Sigma$ that are dependent on the rates of gene switching compared to the degradation rates of the mRNAs. 

These eigenvalues represent the inverse of the relaxation time scales of the system, and are in correspondence with results in the fast-switching (marginal single gene state) limit found in \cite{jia2018relaxation}. This almost completes the solution for the generating function of the telegraph model, aside from determination of the constants $C_m^1$ and $C_m^2$ which in principle can be done using methods related to Sturm-Liouville theory alongside knowledge of the initial conditions of $G$. For a relevant summary on Sturm-Liouville methods see \cite[Appendix A]{holehouse2022exact}, for the initial conditions of the telegraph model see the Appendix of \cite{iyer2009stochasticity}.

\section{Marginality and coarse-graining of gene expression}\label{sec:red-mods}

\noindent A classic form of model reduction is timescale separation, a type of coarse-graining, which is often employed in gene expression under the `fast gene switching assumption', i.e., the condition that $\mathrm{min}\{\sigma_{\mathrm{off}},\sigma_{\mathrm{on}}\}\gg \mathrm{max}\{\rho_{\mathrm{off}},\rho_{\mathrm{on}},d\}$ \cite{holehouse2019revisiting}. Although there are several ways of employing this via quasi equilibrium or steady state assumptions \cite{haseltine2002approximate,rao2003stochastic,kim2015relationship}, the most principled way is via the method of `averaging' \cite{bo2017multiple,jia2020dynamical}. Experimental evidence for fast gene switching can be found in \cite{sepulveda2016measurement} among other studies. Denoting $f=\sigma_{\mathrm{off}}/(\sigma_{\mathrm{off}}+\sigma_{\mathrm{on}})$, the method of averaging then gives the following timescale reduced model valid under the fast gene switching assumption,
\begin{align}
    G\xrightarrow{\rho}G+M,\; M\xrightarrow{d}\varnothing,
\end{align}
where $\rho=f\rho_{\mathrm{off}}+(1-f)\rho_{\mathrm{on}}$, which is nothing other than a one-dimensional microscopically reversible Markov chain which satisfies detailed balance and hence has zero entropy production in the steady state \cite{gardiner1985handbook}. However, taking the same fast switching limit of Eq.~\eqref{eq:entropy-prod-rate} reveals,
$$\dot{S}_{\mathrm{fast}} = (\rho_{\mathrm{on}}-\rho_{\mathrm{off}})f(1-f)\ln\left( \frac{\rho_{\mathrm{on}}}{\rho_{\mathrm{off}}} \right),$$
which is generally non-zero. This trivial example clearly shows that although common methods of model reduction may capture the dynamics of stochastic gene expression, they lose thermodynamic information, a point made in numerous other studies \cite{jia2016model,bo2014entropy}. Notably, as stated in \cite{bo2014entropy},
\begin{displayquote}
    \textit{The limiting entropy production, for arbitrarily large time-scale separation, does not coincide with the entropy production of the effective process.}
\end{displayquote}
Even in the case where $\rho_{\mathrm{off}}=0$ and the dynamics are irreversible, the reduced model still satisfies detailed balance and admits an equilibrium. 

\begin{figure}[ht]
    \includegraphics[width=.3\textwidth]{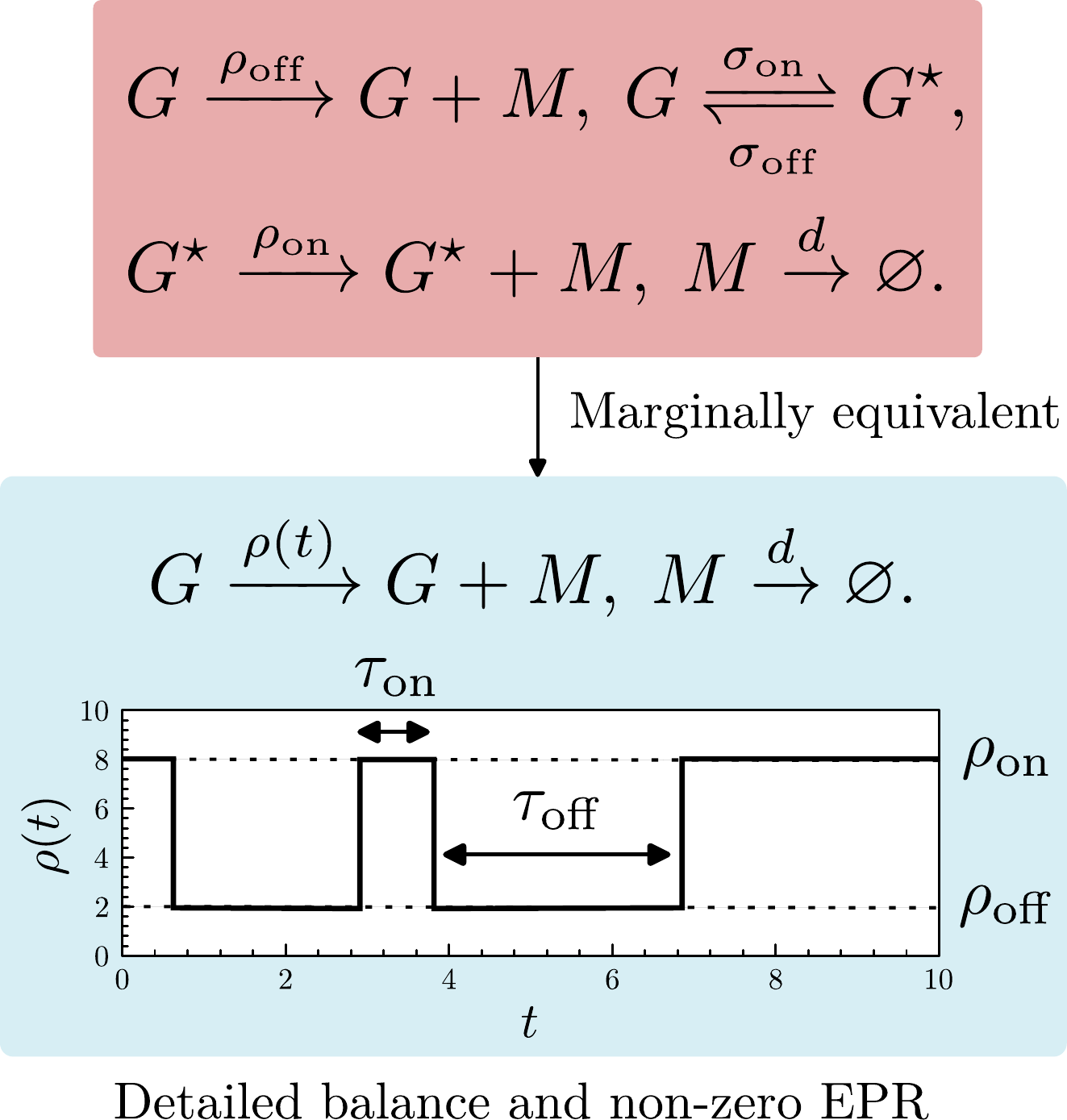}
    \caption{\textbf{Equivalency between the two-state model and a birth-death process with a fluctuating production rate.} Dwell times $\tau_b$ and $\tau_u$ are respectively drawn from the exponential distributions $\sigma_{\mathrm{off}} e^{-\sigma_{\mathrm{off}} \tau}$ and $\sigma_{\mathrm{on}} e^{-\sigma_{\mathrm{on}} \tau}$.}
    \label{fig:fig4}
\end{figure}

A more salient aspect of the analysis presented herein regards the interpretation of the marginal detailed balance without timescale separation arguments. Experiments often only give insight into the state of mRNA or protein numbers, and rarely into the gene states that are theoretically constructed. Hence, if all that can be observed is the mRNA dynamics, the two-state model is equivalent to a birth-death process with a time-dependent mRNA production rate $\rho(t)$, as shown in Fig.~\ref{fig:fig4}. Although the equivalent system has a noisy time-dependent rate (whose statistics are specified in the caption), it is still Markovian, and it provides an example of non-zero entropy production in a system that satisfies detailed balance. That is, \textit{a case of detailed balance with non-zero EPR}. The caveat here is that although the dynamics are Markovian, the time-dependence of $\rho(t)$ allows for this seemingly contradictory situation. When $\rho_{\mathrm{off}}=0$ one can even have detailed balance and an infinite entropy production rate. This provides a striking departure from the standard narrative that detailed balance is reserved for systems in contact with a single reservoir of energy and/or particles or multi-component closed systems \cite{van1976equilibrium,van1992stochastic}. A key takeaway is that although detailed-balance and zero EPR are often stated as being synonymous, this is not necessarily true when the time-independence of the kinetic rates is violated, as can be the case for systems with hidden Markov states.

\section{Calculating the entropy production rate}\label{sec:EPRcalc}

\noindent Following Schnakenberg \cite[Eqs.~(7.6)-(7.11)]{schnakenberg1976network}, the entropy production rate (EPR) is given by,
\begin{align}
    \dot{s} = \frac{k_B}{2}\sum_{\mathbf{x},\mathbf{x'}} J(\mathbf{x'}\to\mathbf{x})\ln\left(  \frac{w(\mathbf{x'}\to\mathbf{x})P(\mathbf{x'},t)}{w(\mathbf{x}\to\mathbf{x'})P(\mathbf{x},t)}\right).
\end{align}
The EPR can be split into two contributions, the first corresponding to the entropy change in the system, the second corresponding to the entropy flowing from the reservoir into the system (i.e., that resulting from coupling the system of interest to external forces). One can write this as,
\begin{align}
    \dot{s} = \frac{k_B}{2}\sum_{\mathbf{x},\mathbf{x'}} J(\mathbf{x'}\to\mathbf{x})\ln\left(  \frac{P(\mathbf{x'},t)}{P(\mathbf{x},t)}\right) + \frac{k_B}{2}\sum_{\mathbf{x},\mathbf{x'}} J(\mathbf{x'}\to\mathbf{x})\ln\left(  \frac{w(\mathbf{x'}\to\mathbf{x}))}{w(\mathbf{x}\to\mathbf{x'})}\right).
\end{align}
The first term on the right-hand side is simply $\mathrm{d}S(t)/\mathrm{d}t$, where $S(t)=-\sum_{\mathbf{x}}P(\mathbf{x},t)\ln(P(\mathbf{x},t))$, i.e., the derivative of the entropy change of the system. The second term in the sum is then the entropy change due to coupling with external reservoirs, typically known as the entropy flow---the literature commonly defines this as the negative of the entropy flow rate, since one considers work done by the system, not on the system. In the case where the system is in a non-equilibrium steady state, then the entropy change in the system becomes zero, i.e., $\mathrm{d}S(t)/\mathrm{d}t=0$, and the EPR becomes the negative of the entropy flow rate \cite{esposito2010three,boeger2022energetics,kirchberg2023energy},
\begin{align}
    \dot{s} = - \frac{k_B}{2}\sum_{\mathbf{x},\mathbf{x'}} J(\mathbf{x'}\to\mathbf{x})\ln\left(  \frac{w(\mathbf{x}\to\mathbf{x'}))}{w(\mathbf{x'}\to\mathbf{x})}\right).
\end{align}
Once the connected states have been identified, we arrive at Eq.~\eqref{eq:macEPR} in the main text. In Eq.~\eqref{eq:macEPR} there are 2 sums that need to be evaluated. The first is $\sum_{n=1}^\infty a_n$ which simply evaluates to $-b_1$ since $a_n = b_{n+1}-b_n$. One also needs to evaluate the sum $\mathcal{S} = \sum_{n=1}^\infty b_n$, where, for reference, $b_n$ is given by
\begin{align}\label{eq:bdef}
    b_n \equiv n d P_0(n) - \rho_{\mathrm{off}}  P_0(n-1).
\end{align}
The sum $\mathcal{S}$ can then be calculated by using $G_0(z)$ from the main text. Multiplying Eq.~\eqref{eq:bdef} by $z^n$ and summing over all $n$ gives,
\begin{align}
    \sum_n b_n z^n = d z G_0'(z) - \rho_{\mathrm{off}}  z G_0(z).
\end{align}
Evaluating this at $z=1$, using $G_0(z)$ from the main text, can be shown to give,
\begin{align}
    \mathcal{S} = \frac{(\rho_{\mathrm{on}} -\rho_{\mathrm{off}} )\sigma_{\mathrm{on}}  \sigma_{\mathrm{off}} }{(\sigma_{\mathrm{on}} +\sigma_{\mathrm{off}} )(d+\sigma_{\mathrm{on}} +\sigma_{\mathrm{off}} )},
\end{align}
from which Eq.~\eqref{eq:entropy-prod-rate} follows.

\section{Extrinsic noise expressions}\label{sec:ext_noise_a}
\noindent Following the integral in Eq.~\eqref{eq:extint}, the expressions for the macroscopic EPR in populations with extrinsic noise are
\begin{align}\nonumber
    \dot{s}_{\mathrm{mes}}[\sigma_{\mathrm{off}} ] &= \frac{1}{d}\alpha  \sigma_{\mathrm{on}} e^{\beta  \sigma_{\mathrm{on}}} \left(\rho_{\mathrm{on}}-\rho_{\mathrm{off}}\right) \ln \left(\frac{\rho_{\mathrm{on}} }{\rho_{\mathrm{off}}}\right) \left(E_{\alpha +1}\left(\beta  \sigma_{\mathrm{on}}\right)-e^{\beta  d} E_{\alpha+1}\left(\beta  \left(d+\sigma_{\mathrm{on}}\right)\right)\right),\\\nonumber
    \dot{s}_{\mathrm{mes}}[\rho_{\mathrm{on}} ] &= \frac{\sigma_{\mathrm{on}} \sigma_{\mathrm{off}} \left(-\left(\left(\alpha -\beta  \rho_{\mathrm{off}}\right) \left(\ln (\beta )-\ln\left(\frac{1}{\rho_{\mathrm{off}}}\right)\right)\right)+\psi ^{(0)}(\alpha ) \left(\alpha -\beta  \rho_{\mathrm{off}} \right)+1\right)}{\beta  \left(\sigma_{\mathrm{on}}+\sigma_{\mathrm{off}}\right) \left(\sigma_{\mathrm{on}}+d+\sigma_{\mathrm{off}}\right)},\\\nonumber
    \dot{s}_{\mathrm{mes}}[\rho_{\mathrm{off}} ] &= \frac{\sigma_{\mathrm{on}} \sigma_{\mathrm{off}} \left(\left(\beta  \rho_{\mathrm{on}}-\alpha \right) \ln \left(\beta  \rho_{\mathrm{on}}\right)+\psi ^{(0)}(\alpha ) \left(\alpha -\beta  \rho_{\mathrm{on}}\right)+1\right)}{\beta  \left(\sigma_{\mathrm{on}}+\sigma_{\mathrm{off}} \right) \left(\sigma_{\mathrm{on}} +d+\sigma_{\mathrm{off}}\right)},
\end{align}
where $\psi ^{(0)}(\alpha )$ is a polygamma function defined by $\psi ^{(0)}(\alpha )=\Gamma'(\alpha)/\Gamma(\alpha)$ and again the dependence on the gamma distribution parameters $\alpha$ and $\beta$ has been dropped for brevity. The corresponding error functions, as defined in Eq.~\eqref{eq:errorfn}, are given by,
\begin{align}\nonumber
    e_{\sigma_{\mathrm{off}} } &= \frac{1}{\beta  d}e^{\beta  \sigma_{\mathrm{on}}} \left(\alpha +\beta  \sigma_{\mathrm{on}}\right) \left(\alpha +\beta  \left(d+\sigma_{\mathrm{on}}\right)\right) \left(E_{\alpha+1}\left(\beta  \sigma_{\mathrm{on}}\right)-e^{\beta  d} E_{\alpha +1}\left(\beta  \left(d+\sigma_{\mathrm{on}} \right)\right)\right)-1,\\\nonumber
    e_{\rho_{\mathrm{on}} } &= \frac{\psi ^{(0)}(\alpha ) \left(\alpha -\beta  \rho_{\mathrm{off}} \right)-\left(\alpha -\beta  \rho_{\mathrm{off}}\right) \left(\ln (\beta )+\ln \left(\frac{\alpha }{\beta  \rho_{\mathrm{off}}}\right)+\ln \left(\rho_{\mathrm{off}}\right)\right)+1}{\left(\alpha -\beta  \rho_{\mathrm{off}}\right) \ln \left(\frac{\alpha }{\beta  \rho_{\mathrm{off}}}\right)},\\\nonumber
    e_{\rho_{\mathrm{off}} } &= \frac{-\psi ^{(0)}(\alpha )+\frac{1}{\beta  \rho_{\mathrm{on}}-\alpha }-\ln \left(\frac{\beta  \rho_{\mathrm{on}}}{\alpha}\right)+\ln \left(\beta  \rho_{\mathrm{on}}\right)}{\ln \left(\frac{\beta  \rho_{\mathrm{on}}}{\alpha }\right)}.
\end{align}

\clearpage
\section{Supplementary Figures}

\begin{figure}[h!]
    \centering
    \includegraphics[width=.6\textwidth]{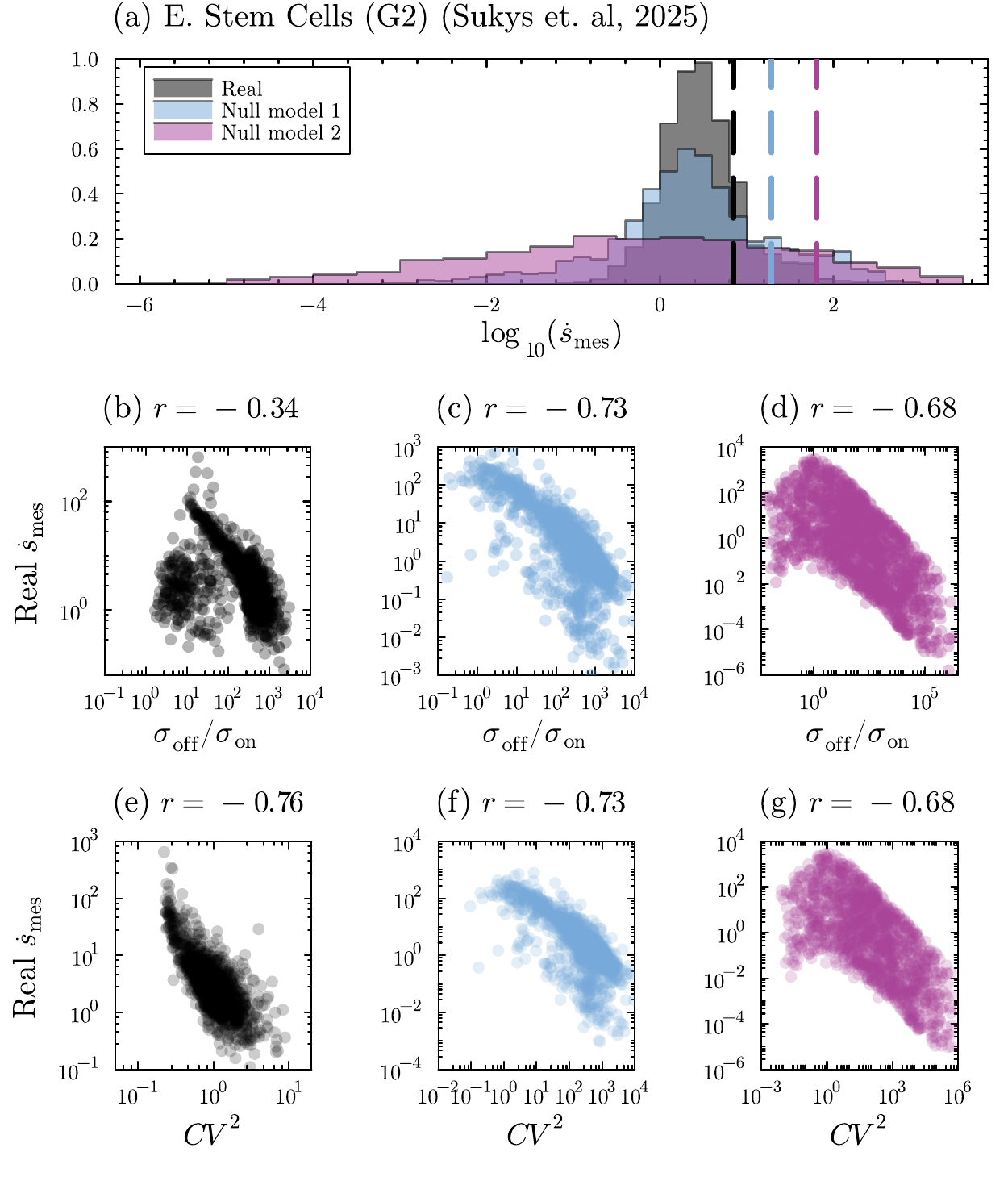}
    \caption{\textbf{Analog of Fig.~\ref{fig:fig3} but for 1,436 G2M cell-cycle stage mouse fibroblast genes \cite{sukys2025cell}.} }
    \label{fig:sukysg2m}
\end{figure}

\begin{figure}[h!]
    \centering
    \includegraphics[width=.6\textwidth]{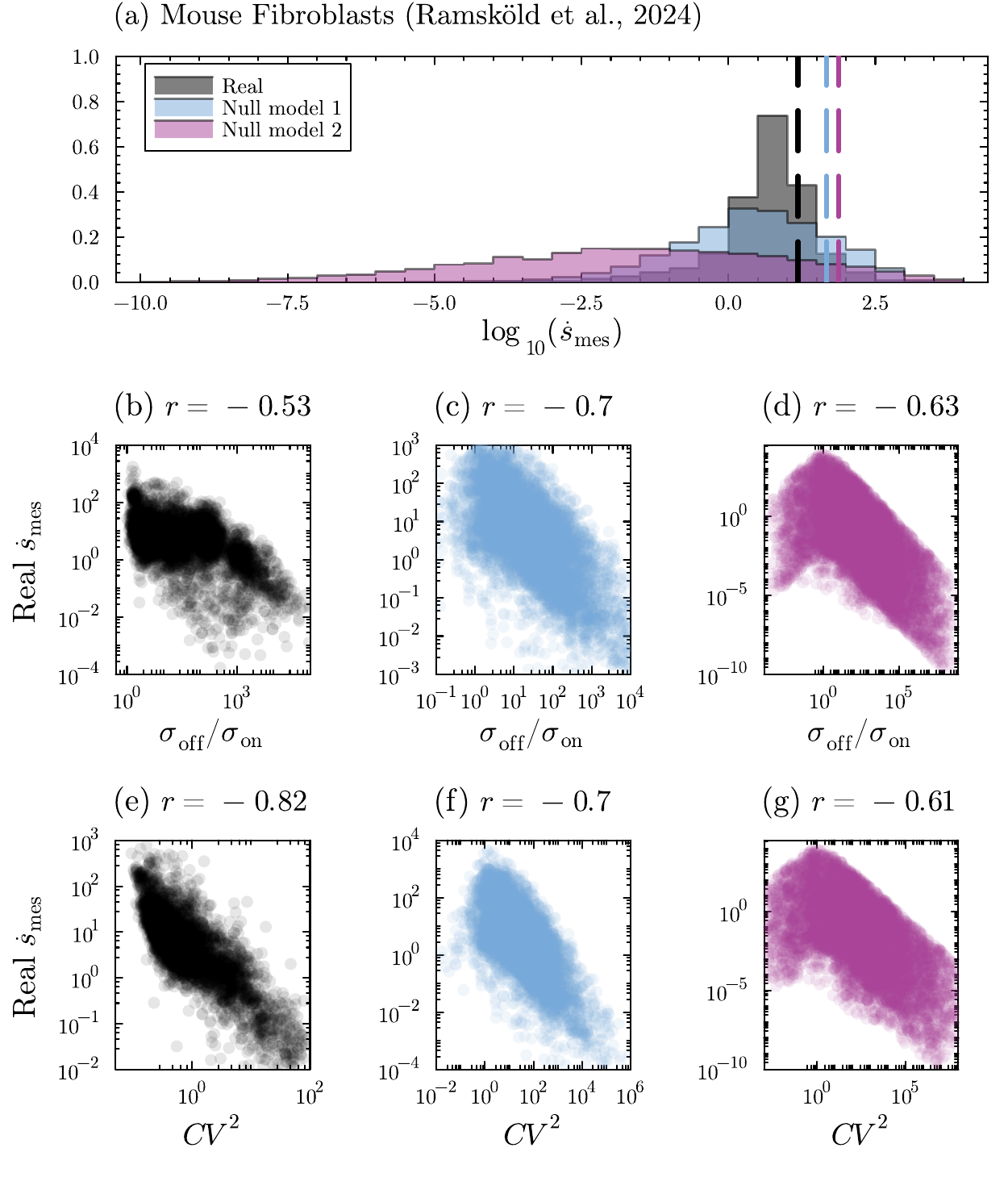}
    \caption{\textbf{Analog of Fig.~\ref{fig:fig3} but for 5,623 mouse fibroblast genes from \cite{ramskold2024single}.} }
    \label{fig:2024ramskold}
\end{figure}

\begin{figure}[h!]
    \centering
    \includegraphics[width=.6\textwidth]{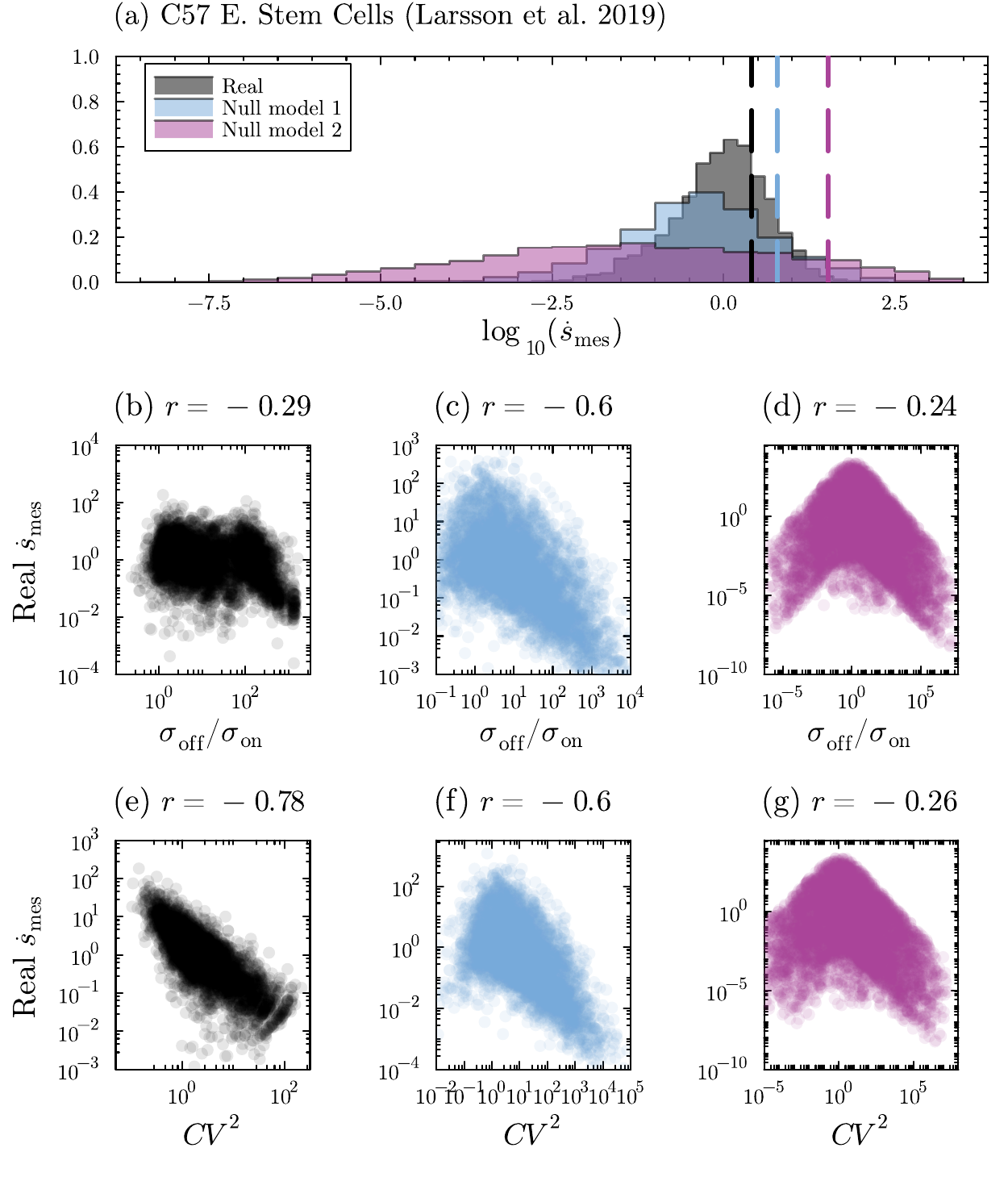}
    \caption{\textbf{Analog of Fig.~\ref{fig:fig3} but for 5,227 C57 mouse embryonic stem cell genes from \cite{larsson2019genomic}.} }
    \label{fig:2019escc57}
\end{figure}

\begin{figure}[h!]
    \centering
    \includegraphics[width=.6\textwidth]{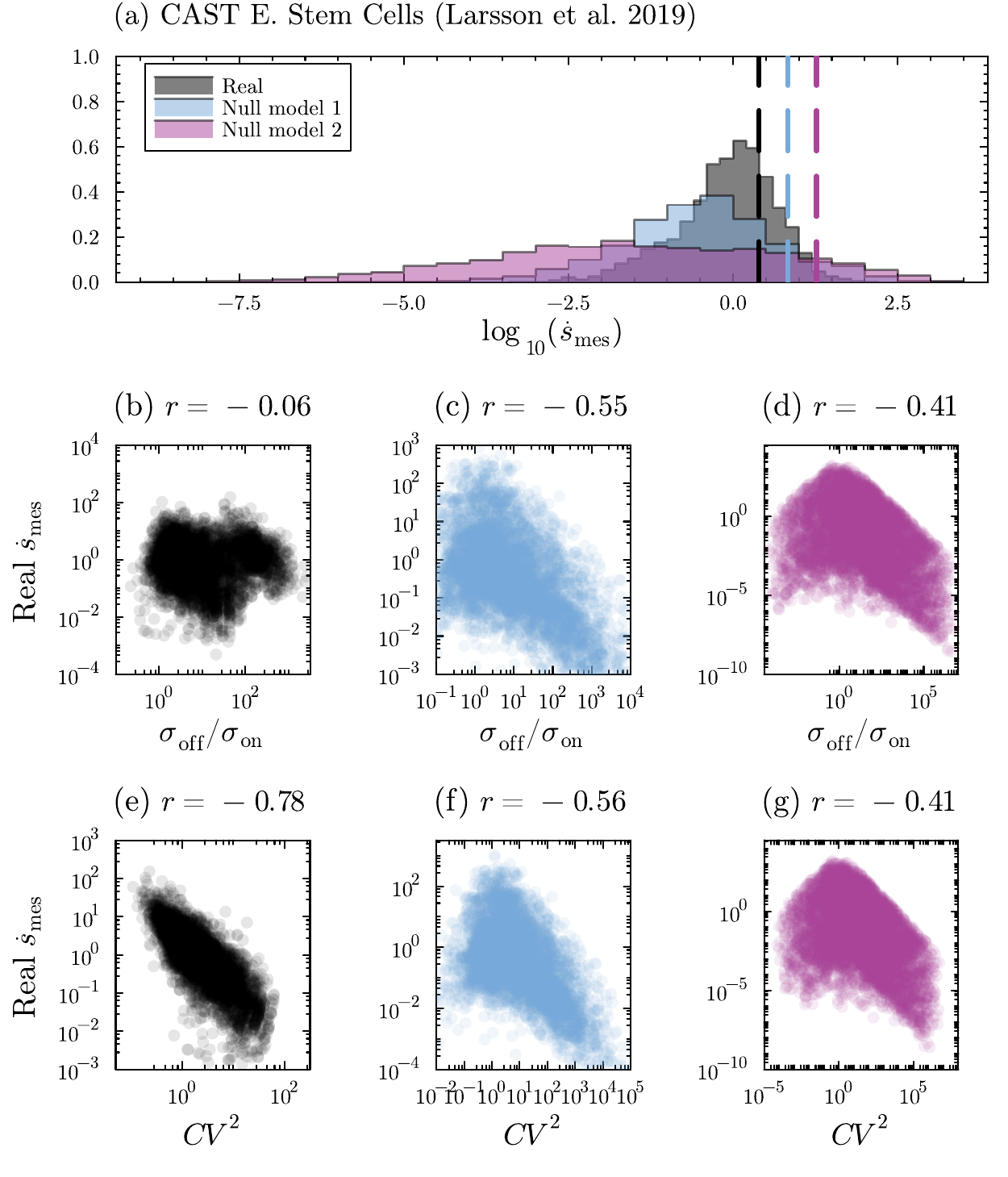}
    \caption{\textbf{Analog of Fig.~\ref{fig:fig3} but for 4,925 CAST mouse embryonic stem cell genes from \cite{larsson2019genomic}.} }
    \label{fig:2019esccast}
\end{figure}

\begin{figure}[h!]
    \centering
    \includegraphics[width=.6\textwidth]{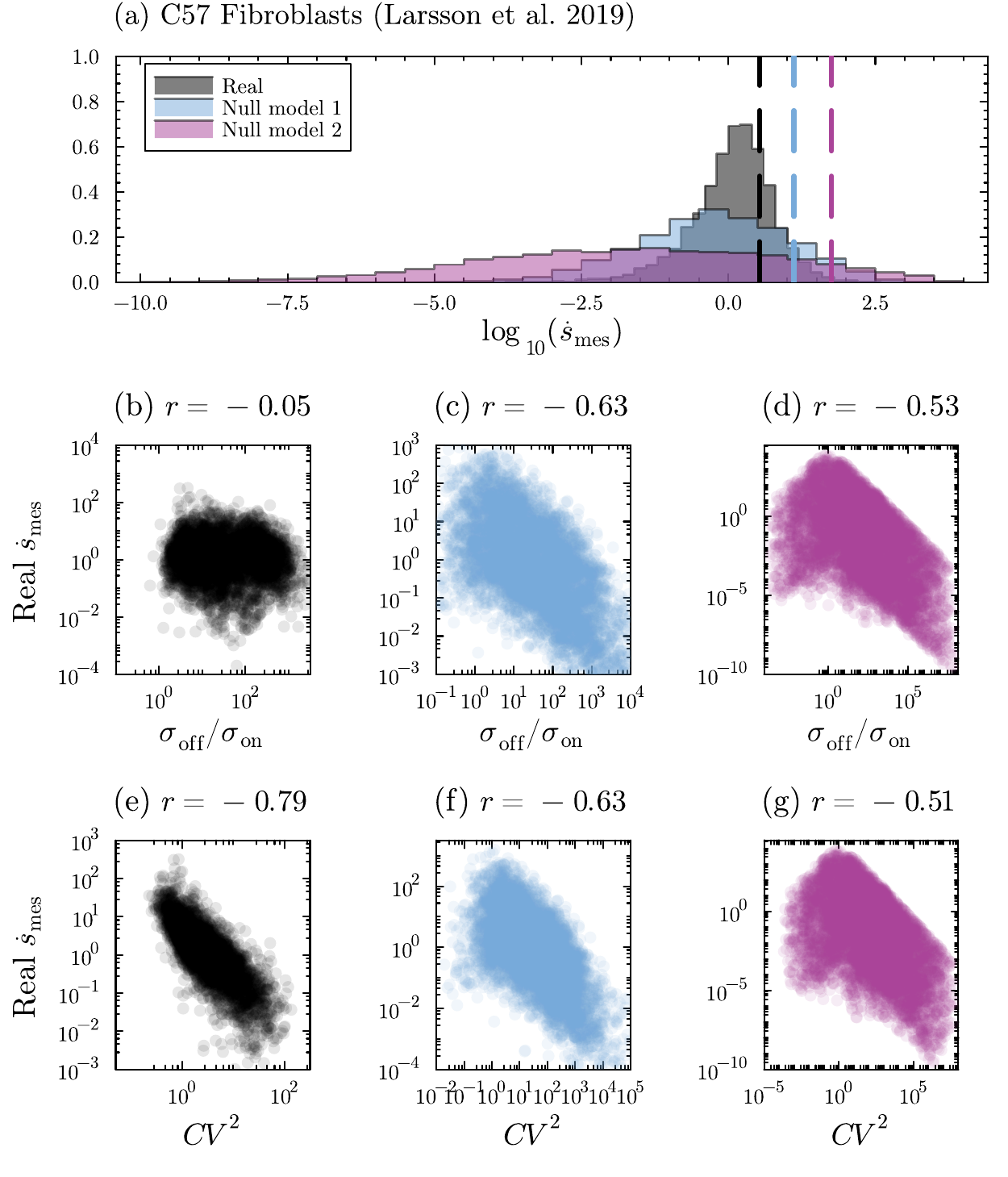}
    \caption{\textbf{Analog of Fig.~\ref{fig:fig3} but for 5,858 C57 mouse fibroblast genes from \cite{larsson2019genomic}.} }
    \label{fig:2019fibc57}
\end{figure}

\begin{figure}[h!]
    \centering
    \includegraphics[width=.6\textwidth]{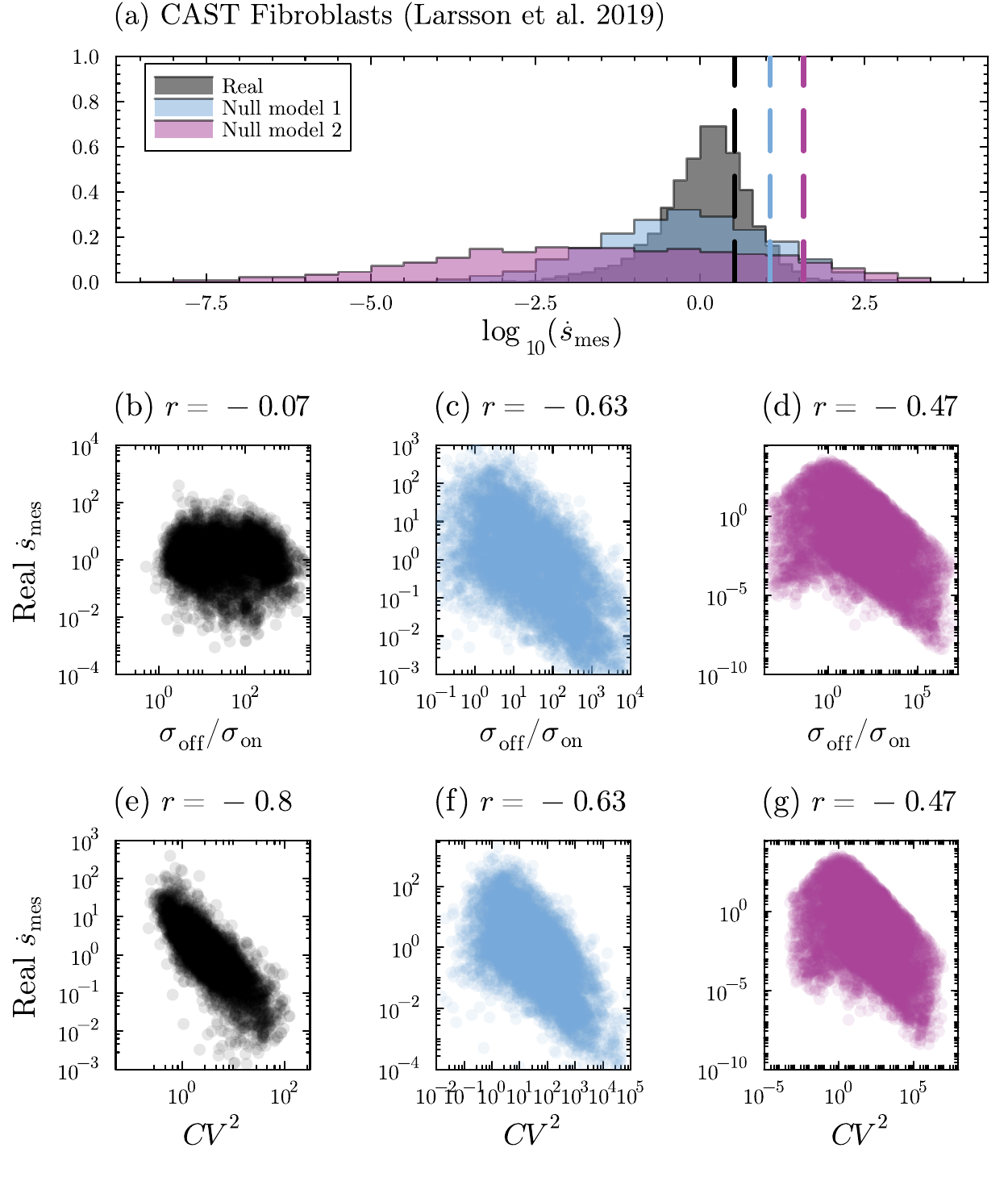}
    \caption{\textbf{Analog of Fig.~\ref{fig:fig3} but for 5,854 CAST mouse fibroblast genes from \cite{larsson2019genomic}.} }
    \label{fig:2019fibcast}
\end{figure}

\begin{figure}[h!]
    \centering
    \includegraphics[width=.3\textwidth]{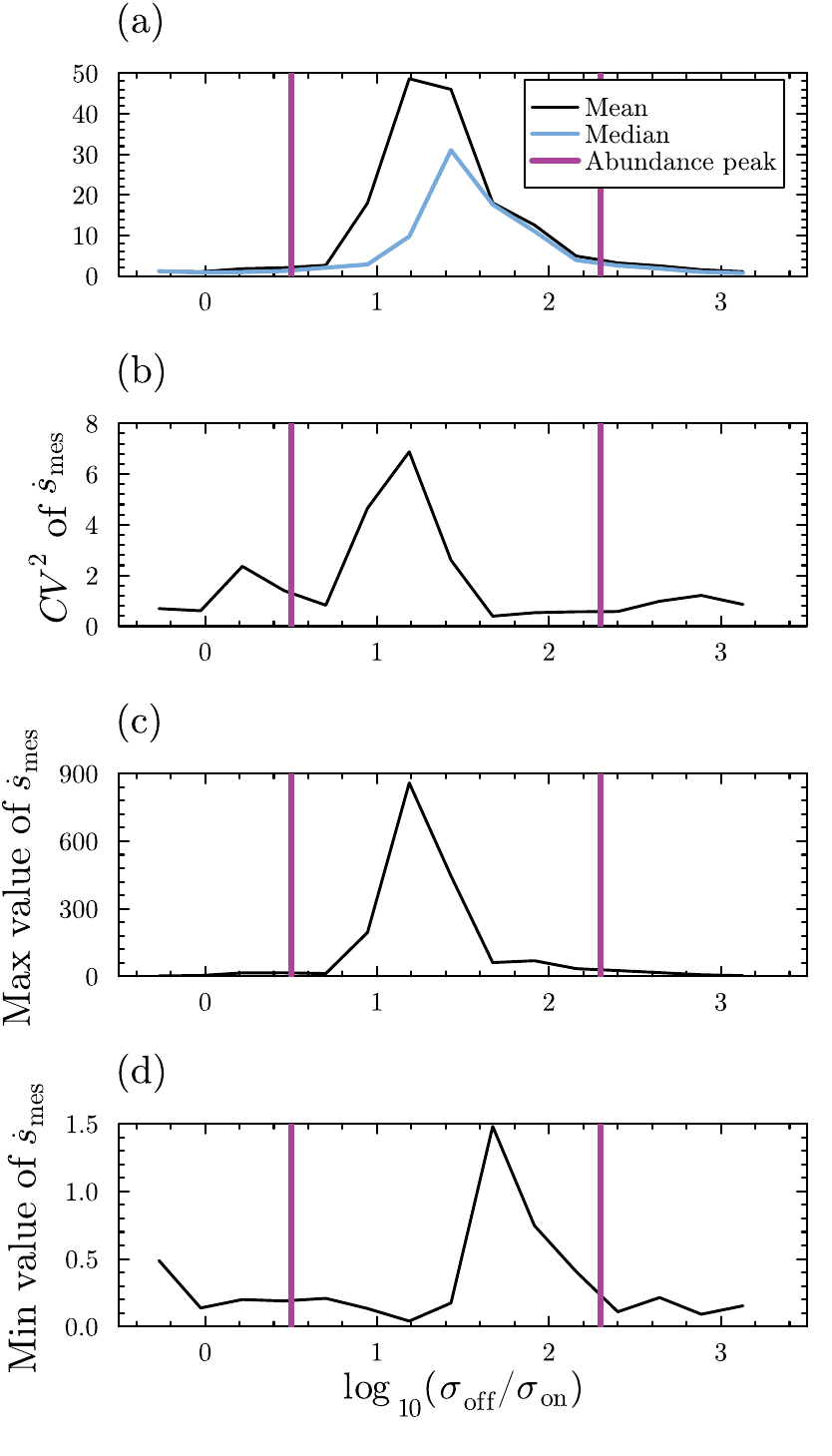}
    \caption{\textbf{Supplementary figure to Fig.~\ref{fig:fig5} showing that the choice of average of $\dot s_{\mathrm{mes}}$ is not important.} (a) Comparison between the median and mean values of binned $\dot s_{\mathrm{mes}}$. (b) Exploring the noise of $\dot s_{\mathrm{mes}}$ in each bin. (c) and (d) show that even if we choose the maximum or minimum values of $\dot s_{\mathrm{mes}}$ in each bin then the peak of $\dot s_{\mathrm{mes}}$ still resides between the abundance peaks of $\log_{10}(\sigma_{\mathrm{off}}/\sigma_{\mathrm{on}})$. }
    \label{fig:medminmax}
\end{figure}

\begin{figure}[h!]
    \centering
    \includegraphics[width=.6\textwidth]{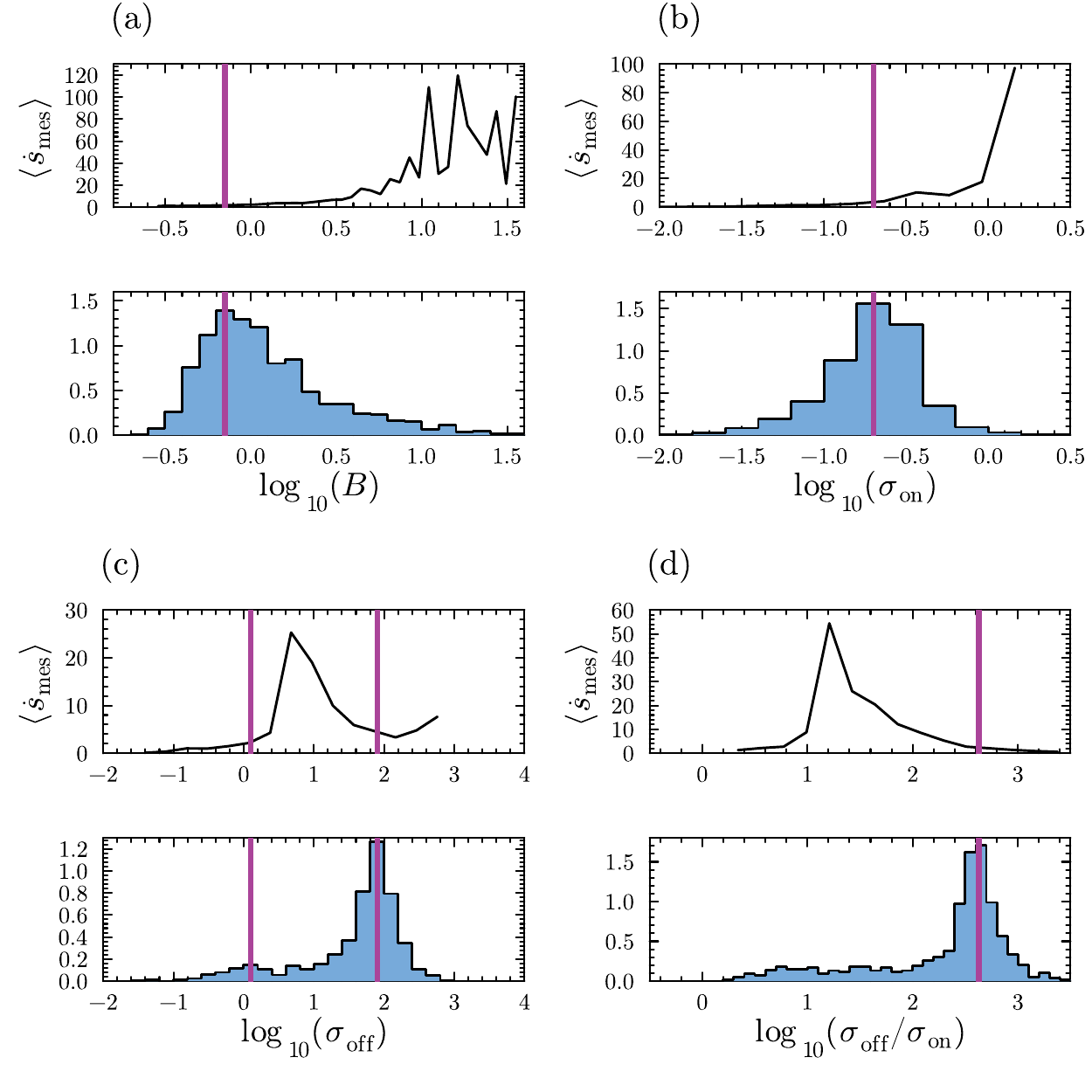}
    \caption{\textbf{Analog of Fig.~\ref{fig:fig5} but for G2M cell-cycle stage mouse fibroblasts \cite{sukys2025cell}.} }
    \label{fig:bmsg2}
\end{figure}

\begin{figure}[h!]
    \centering
    \includegraphics[width=.6\textwidth]{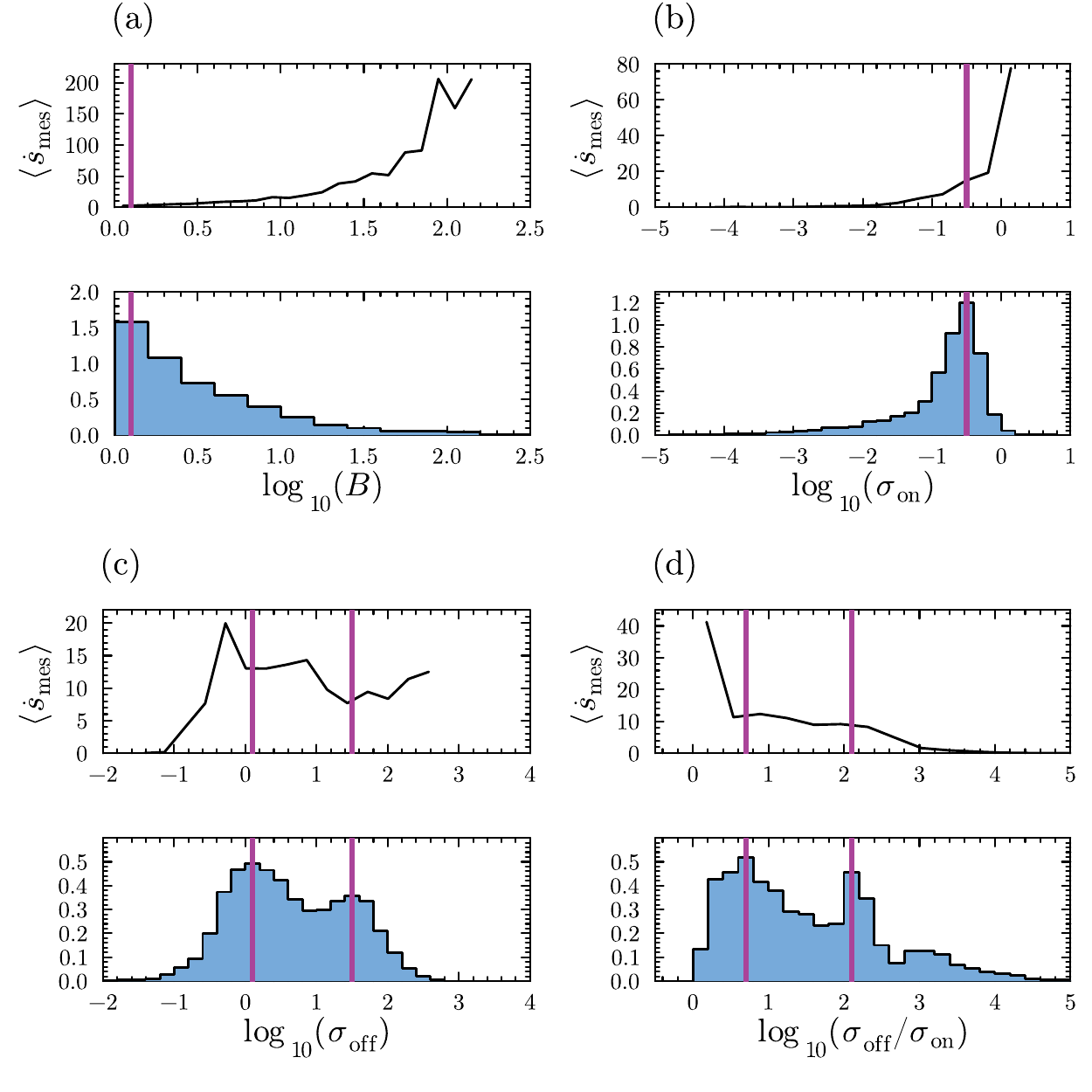}
    \caption{\textbf{Analog of Fig.~\ref{fig:fig5} but for mouse fibroblast cells from \cite{ramskold2024single}.} }
    \label{fig:bmsr}
\end{figure}

\begin{figure}[h!]
    \centering
    \includegraphics[width=.6\textwidth]{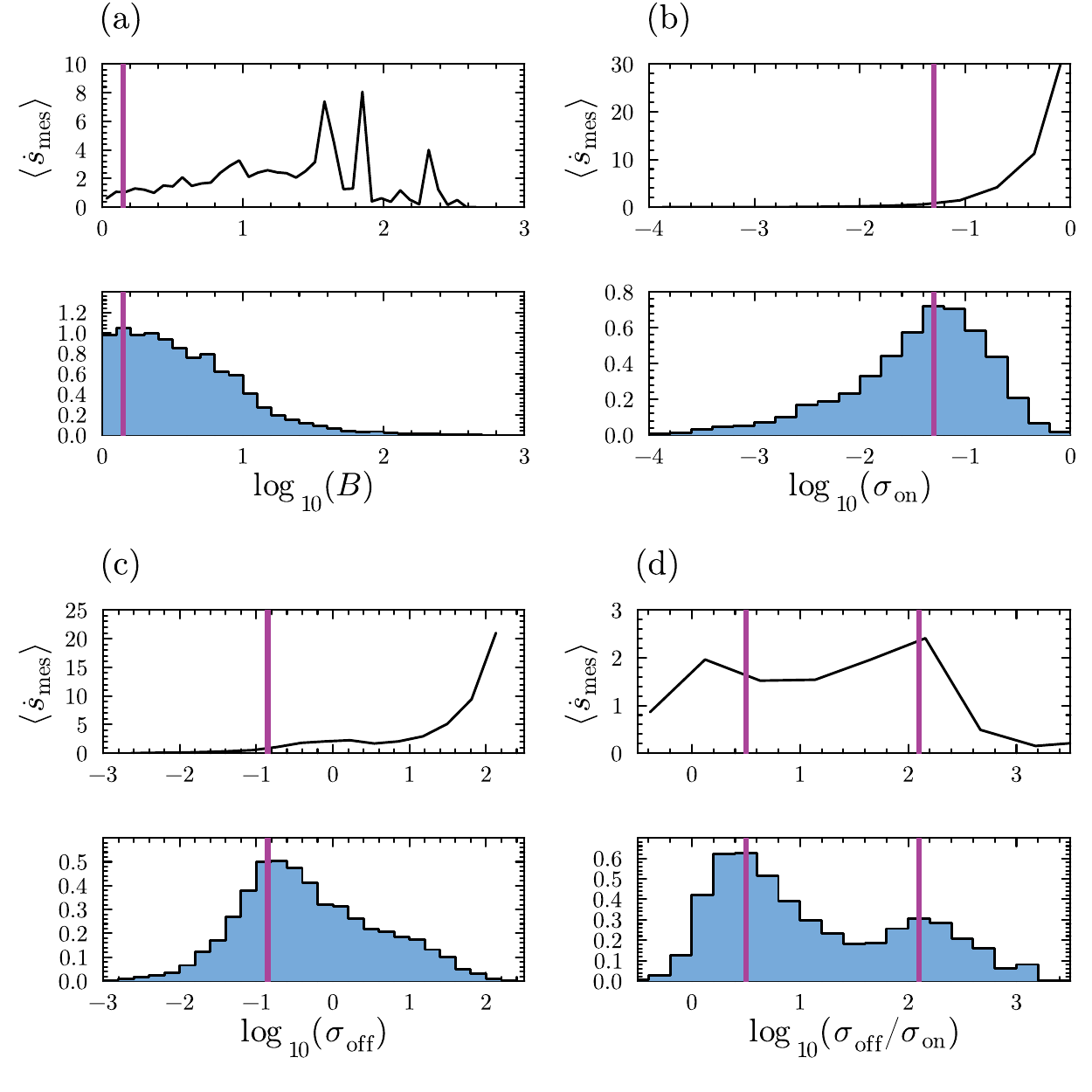}
    \caption{\textbf{Analog of Fig.~\ref{fig:fig5} but for C57 mouse embryonic stem cells from \cite{larsson2019genomic}.} }
    \label{fig:bms2019escc57}
\end{figure}

\begin{figure}[h!]
    \centering
    \includegraphics[width=.6\textwidth]{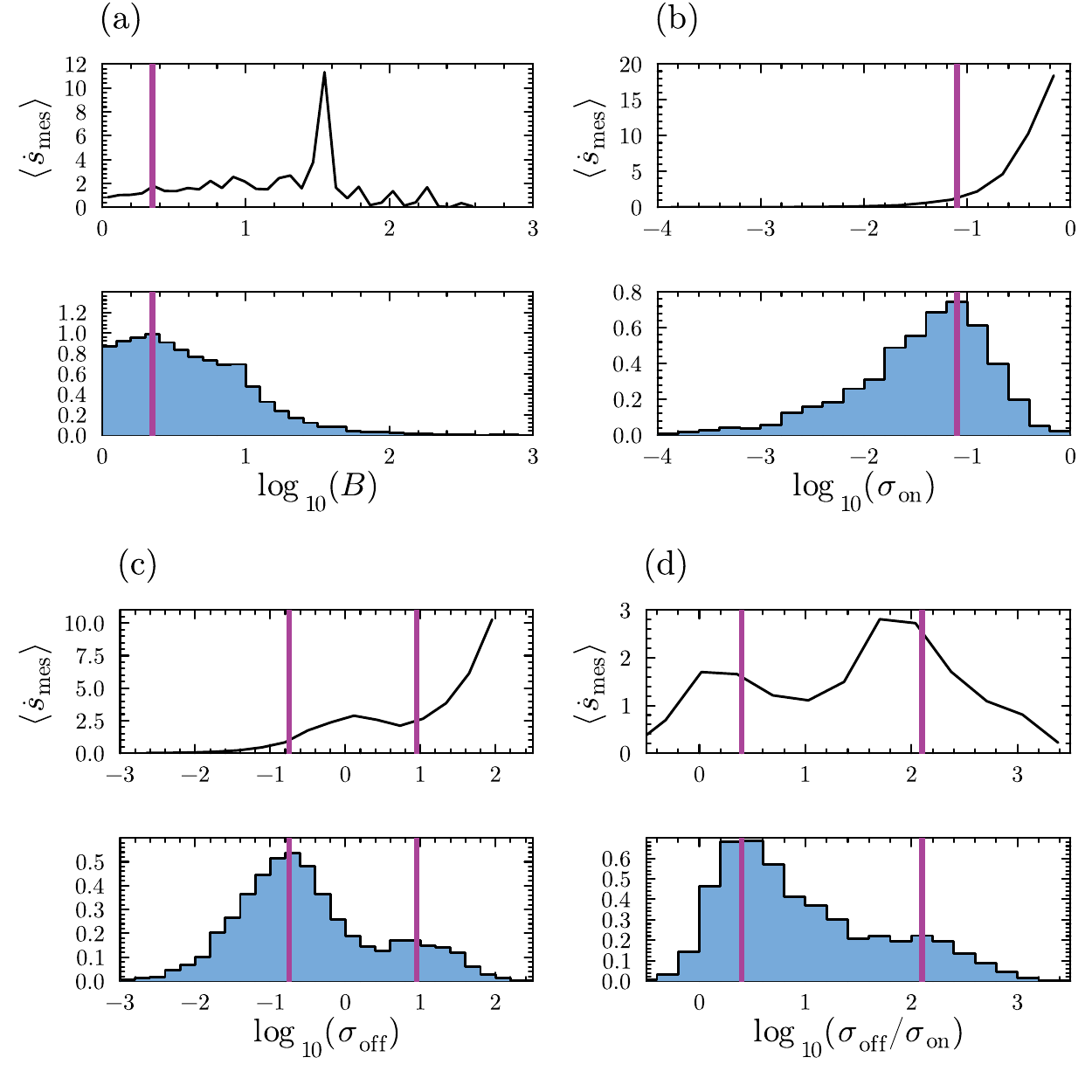}
    \caption{\textbf{Analog of Fig.~\ref{fig:fig5} but for CAST mouse embryonic stem cells from \cite{larsson2019genomic}.} }
    \label{fig:bms2019esccast}
\end{figure}

\begin{figure}[h!]
    \centering
    \includegraphics[width=.6\textwidth]{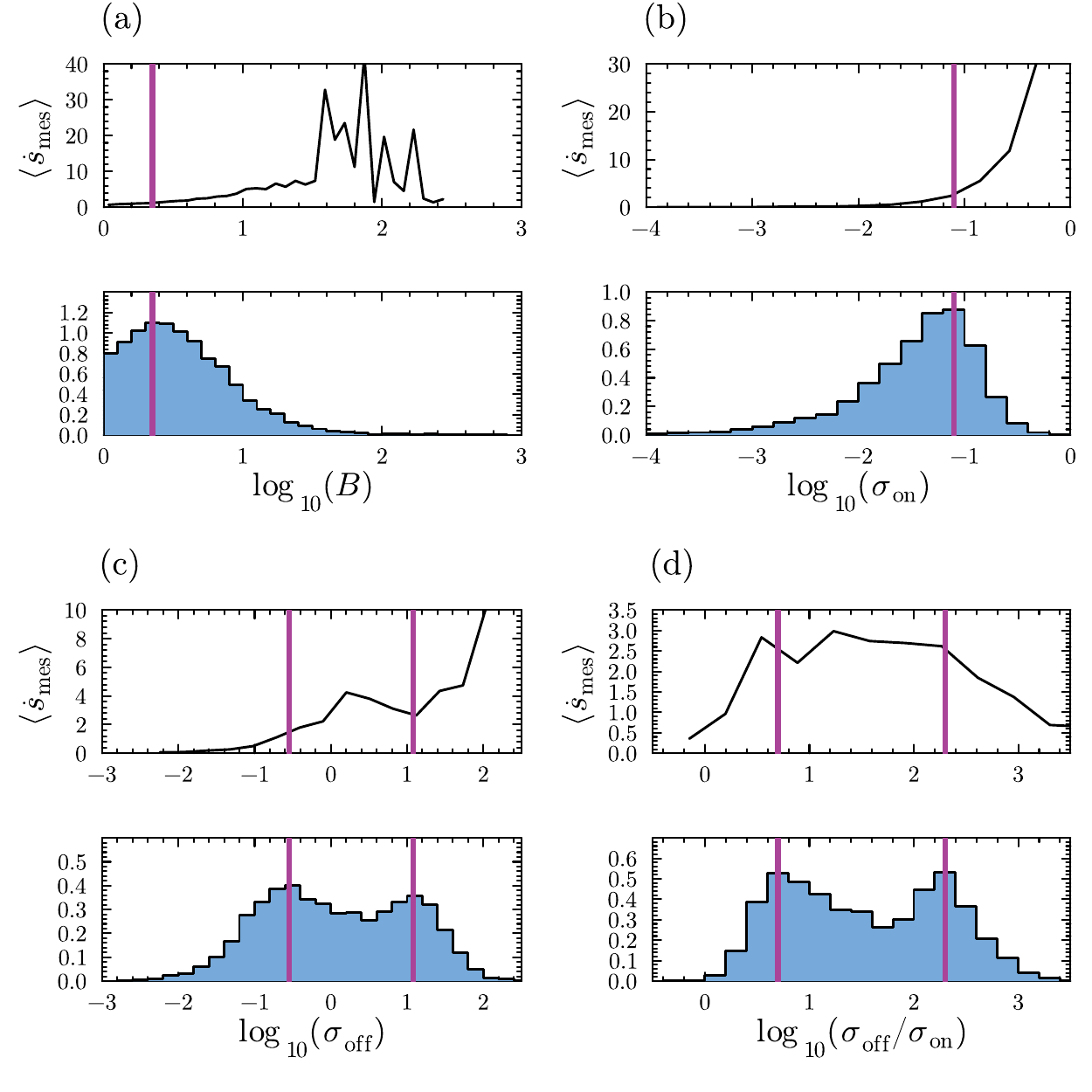}
    \caption{\textbf{Analog of Fig.~\ref{fig:fig5} but for C57 mouse fibroblast cells from \cite{larsson2019genomic}.} }
    \label{fig:bms2019fc57}
\end{figure}

\begin{figure}[h!]
    \centering
    \includegraphics[width=.6\textwidth]{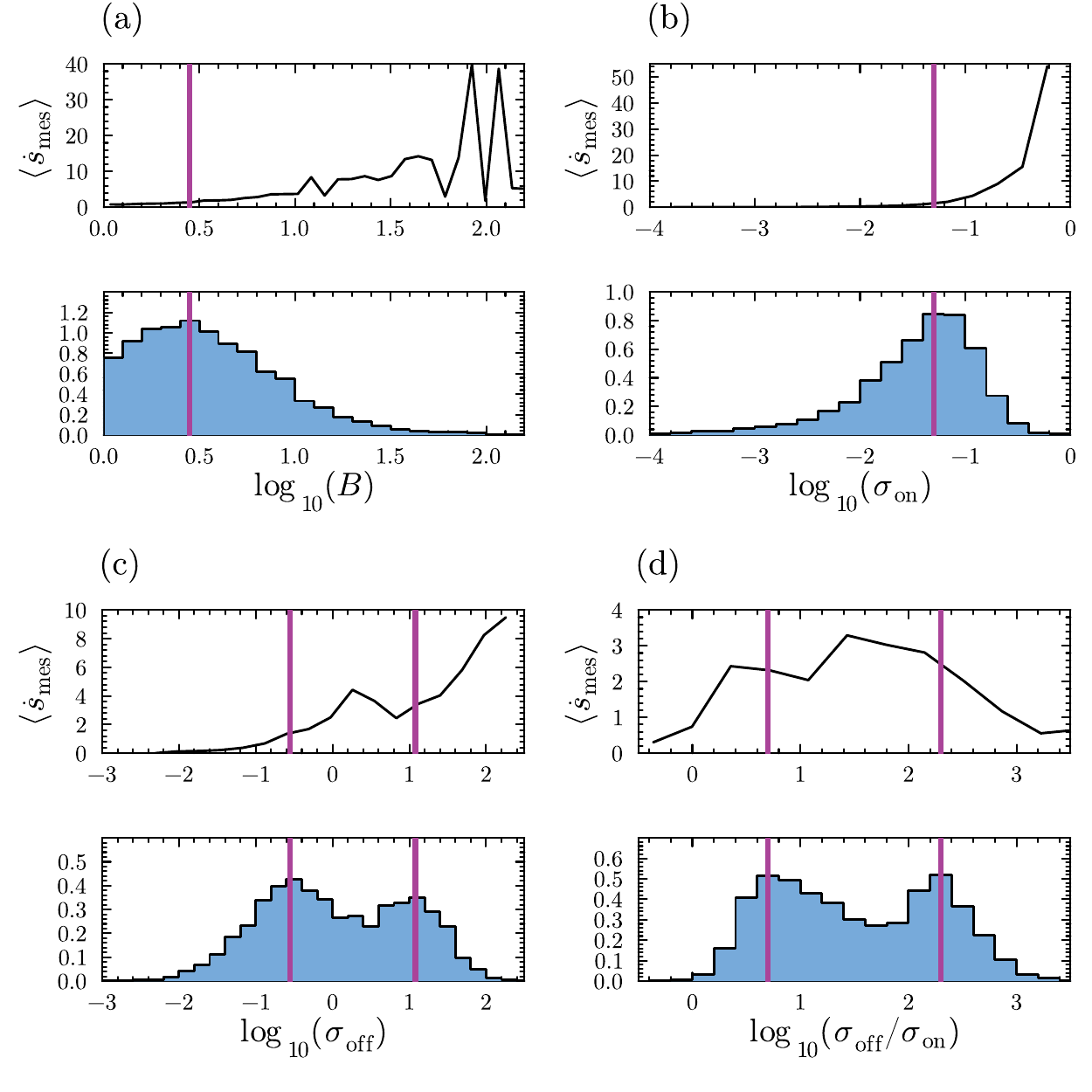}
    \caption{\textbf{Analog of Fig.~\ref{fig:fig5} but for CAST mouse fibroblast cells from \cite{larsson2019genomic}.} }
    \label{fig:bms2019fcast}
\end{figure}

\begin{figure}[h!]
    \centering
    \includegraphics[width=.6\textwidth]{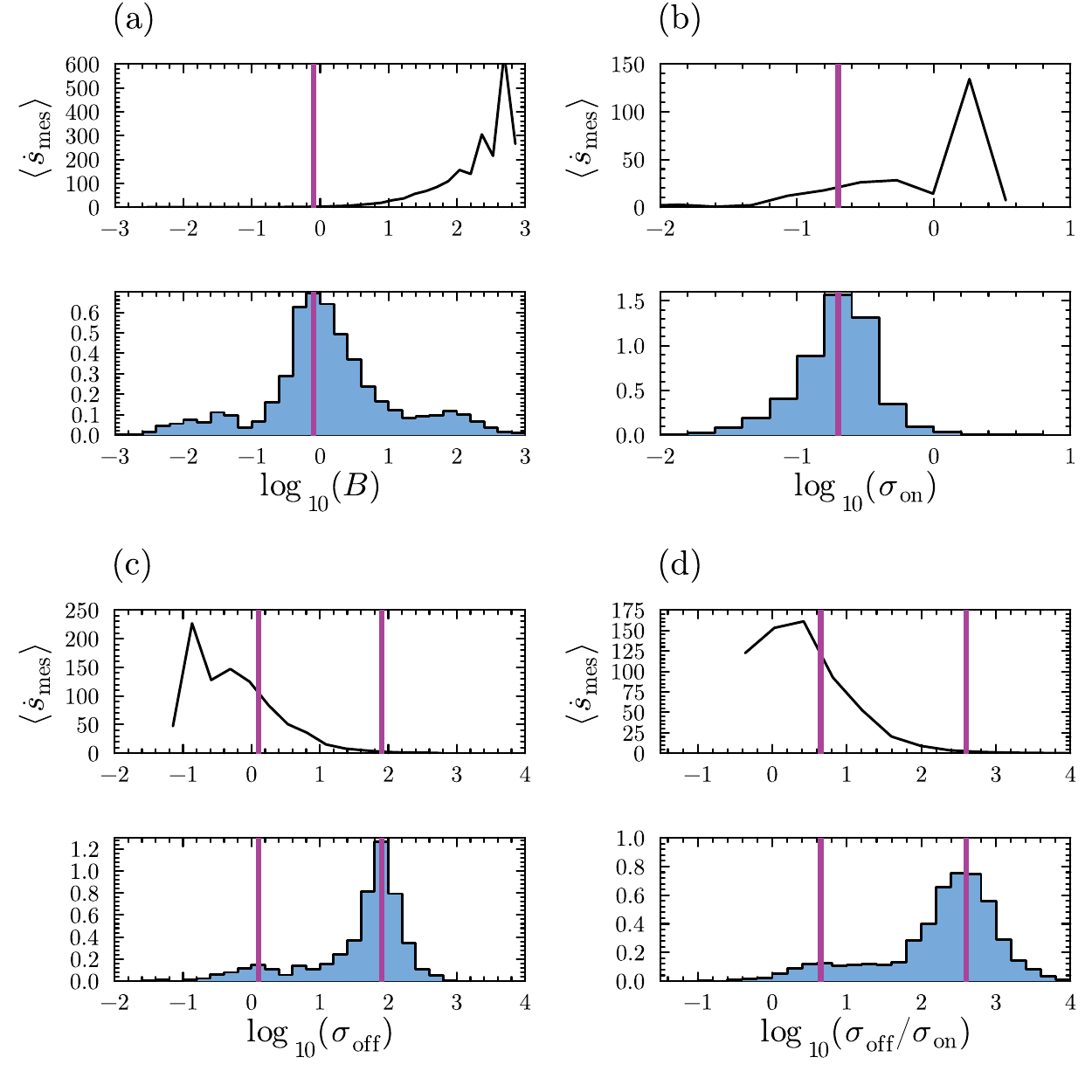}
    \caption{\textbf{Analog of Fig.~\ref{fig:fig-null-bin1} but for G2M cell-cycle stage mouse fibroblasts \cite{sukys2025cell}.} }
    \label{fig:g2mnull1}
\end{figure}

\begin{figure}[h!]
    \centering
    \includegraphics[width=.6\textwidth]{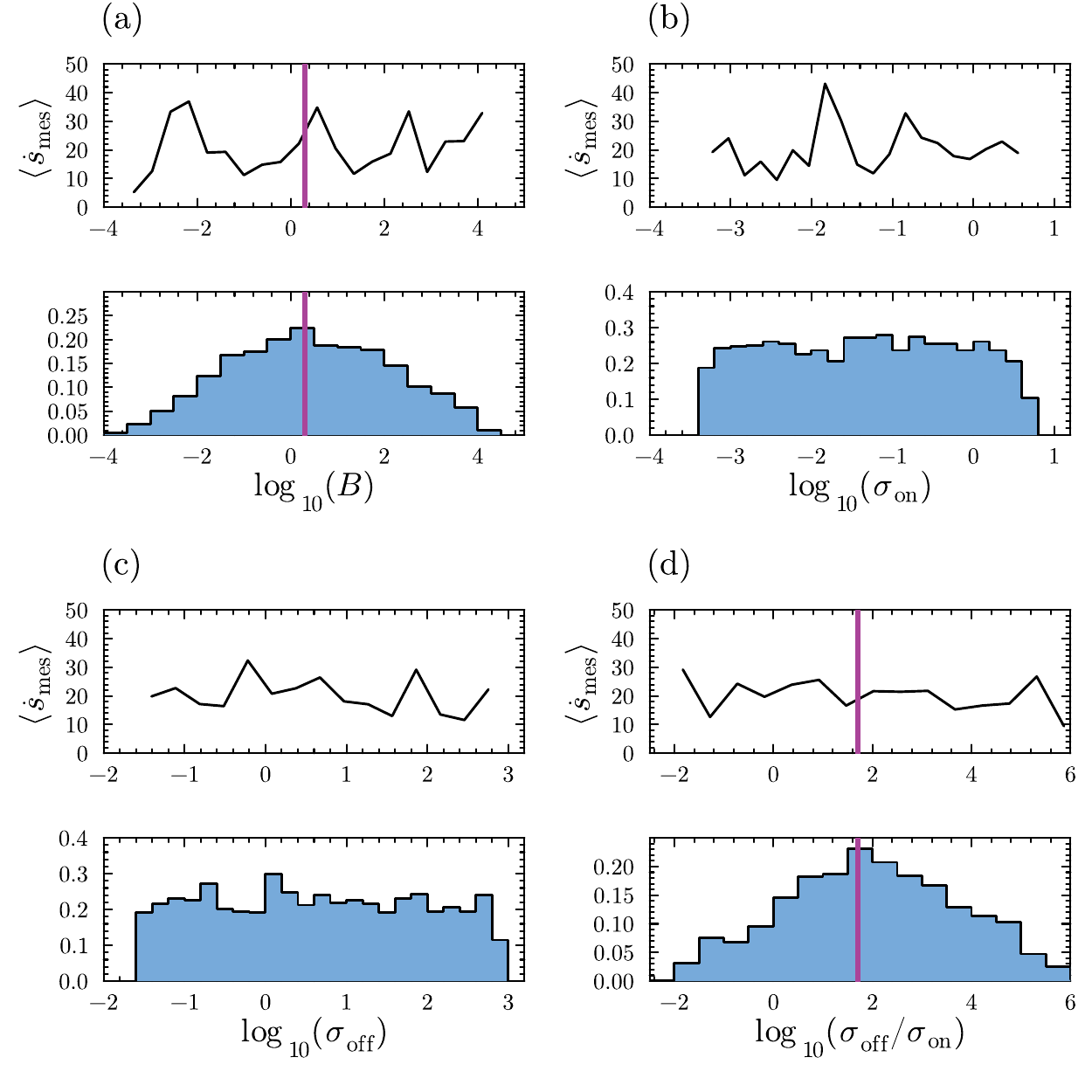}
    \caption{\textbf{Analog of Fig.~\ref{fig:fig-null-bin2} but for G2M cell-cycle stage mouse fibroblasts \cite{sukys2025cell}.} }
    \label{fig:g2mnull2}
\end{figure}

\end{document}